\documentclass[aps,pra,notitlepage,superscriptaddress,showpacs]{revtex4-2}
\usepackage[T1]{fontenc}
\usepackage{amsmath}
\usepackage{amssymb}
\usepackage{graphicx}
\usepackage{esint}
 
\usepackage{hyperref}
\hypersetup{colorlinks=true,citecolor=blue,linkcolor=red,urlcolor=blue}
\begin{document}
\title{Spatio-spectral control of spontaneous emission }
\author{Seyyed Hossein Asadpour}
\email{asadpour@ipm.ir}

\affiliation{School of Physics, Institute for Research in Fundamental Sciences (IPM), Tehran 19395-5531, Iran}
\author{Muqaddar Abbas}
\email{muqaddarabbas@xjtu.edu.cn}
\affiliation{Ministry of Education Key Laboratory for Nonequilibrium Synthesis
and Modulation of Condensed Matter, Shaanxi Province Key Laboratory
of Quantum Information and Quantum Optoelectronic Devices, School
of Physics, Xi\textquoteright an Jiaotong University, Xi\textquoteright an
710049, China}
\author{Hamid R. Hamedi}
\email{hamid.hamedi@tfai.vu.lt}

\affiliation{Institute of Theoretical Physics and Astronomy, Vilnius University,
Sauletekio 3, Vilnius 10257, Lithuania}
\author{Julius Ruseckas}
\email{julius.ruseckas@gmail.com }

\affiliation{Baltic Institute of Advanced Technology, 01124 Vilnius, Lithuania}
\author{Emmanuel Paspalakis}
\email{paspalak@upatras.gr }

\affiliation{Materials Science Department, School of Natural Sciences, University
of Patras, Patras 265 04, Greece}
\author{Reza Asgari}
\email{asgari@ipm.ir}

\affiliation{Department of Physics, Zhejiang Normal University, Jinhua, Zhejiang 321004, China}
\affiliation{School of Physics, Institute for Research in Fundamental Sciences (IPM), Tehran 19395-5531, Iran}
\begin{abstract}
We propose a scheme aimed at achieving spatio-spectral control over spontaneous
emission within a four-level atom-light coupling system interacting with
optical vortices carrying orbital angular momentum (OAM). The atom
comprises a ground level and two excited states coupled with two laser
fields, forming a V subsystem where the upper states exclusively decay
to a common fourth state via two channels. By investigating various initial states of the atom and considering the presence or absence of quantum interference in spontaneous emission channels, we analyze how the characteristics of the OAM-carrying vortex beam imprint onto the emission spectrum. The interplay
between the optical vortex and the quantum system, including its environment
modes, induces a wide variety of spatio-spectral behaviour, including
two-dimensional spectral-peak narrowing, spectral-peak enhancement,
spectral-peak suppression, and spontaneous emission reduction or quenching
in the spatial azimuthal plane. Our findings shed light on the dynamics of atom-vortex beam light interactions and offer insights into the manipulation of emission properties at the quantum level.
\end{abstract}
\maketitle

\section{Introduction}

There has been a notable surge in the study of spontaneous emission
resulting from the interplay between quantum systems and environmental
modes \cite{gerry2023introductory, meystre2021quantum}. Considering various methods and systems aimed at manipulating
and controlling the spectrum of spontaneous emission has evolved into
a captivating frontier within the realm of scientific research. Extensive
efforts have been invested in delving into theoretical frameworks
that address the control of spontaneous emission \citep{ZhouSE96,John,Swain97,Paspalakis98,Ghafoor2000,Ghafoor2002,anton2005,AiJun,Wu2005,Fountoulakis,PaspalakisJMO,JiaHua,Arun,Li08,ChunOE,Thanopulos,Thanopulos2019,Whisler}.
The applications stemming from the control of spontaneous emission
are multifaceted, spanning diverse areas such as lasing without inversion \citep{Agarwal91,Kozlov}, Electromagnetically induced grating
(EIG) \citep{Bin,Forough}, accurate localization and magnetic field
measurement \citep{Scullymagnetometer,Ghafoor2002,Ren,Zhiping}, transparent
materials with high refractive index \citep{Scully91ri,Evangelou},
advanced quantum information processing \citep{Charles,Paternostro}.
The manipulation of spontaneous emission has been demonstrated through
various approaches, including spontaneously generated coherence
\citep{Zhu1996}, relative phase control \citep{Paspalakis98}, control
through incoherent pump processes \citep{Kapale}, external coupling
fields control \citep{Ghafoor2000,JiaHua}, and altering the environmental
conditions of atoms, such as in free-space \citep{Bojer}, in proximity
to plasmonic nanostructures \citep{Evangelou2011}, within photonic
crystals \citep{Zhang2002} and confinement within optical cavities
\citep{Pathak}. Each circumstance presents unique densities of electromagnetic
modes interacting with matter, showcasing the versatility of the control
mechanisms associated with spontaneous emission in quantum systems.

At the same time, optical vortices, known for carrying orbital angular momentum (OAM) \citep{Allen,Arnold},
have recently garnered increased attention due to their promising
applications across various domains. Optical vortices exhibit distinctive characteristics,
featuring helical wavefronts that converge into circular patterns,
deviating from the conventional point focuses. The spotlight has particularly
intensified on these twisted light beams, showcasing their potential
in quantum information processing \citep{Jinwen} biosciences \citep{Stevenson},
microtrapping and alignment \citep{Macdonald}, and optical micromanipulation
\citep{Woerdemann}. However, to comprehensively understand their
impact, it is imperative to delve into the unique nature of the interaction
between optical vortices and matter. The interplay between these unique
structured light beams and matter reveals a multitude of captivating
phenomena and effects, which have been extensively investigated \citep{Dutton,Chen2008,Ruseckas1,Lembessis,Ding,Ruseckas2,Radwell,Sharma,HamidOE1,HamediOFF,Jing,Seyyed,Song,Daloi,HamidOEloc,Hamedimatch,Chen2023,Meng}. 

Recently, optical vortices endowed with OAM
have ushered in a novel realm of possibilities for manipulating electromagnetically
induced transparency (EIT) \citep{Fleischhauer}. This entails the
creation of spatially dependent EIT within a phase-dependent $\Lambda$
scheme \citep{Radwell}, where the phase sensitivity is meticulously
tuned using an external magnetic field. We have proposed a sophisticated
extension to this spatially dependent EIT, replacing the conventional
magnetic field with additional optical transitions \citep{HamidOE1}.
Subsequent studies have revealed that optical vortices exert profound
spatial effects in EIG \citep{Seyyed} and light amplification without inversion  \citep{Hamid2023}.

This paper delves into harnessing the inherent OAM in optical vortices
for precise spatio-spectral control over spontaneous emission in a
four-level atom-light coupling system. The system consists of a ground
level coupled to two excited states connected with two laser fields,
forming a V subsystem. We specifically consider one of the laser beams
as an optical vortex. The upper states decay to a common fourth state
through two decay channels. Various scenarios are explored for the
initial state of the atom, encompassing cases where (i) the atom initiates
from its ground state, (ii) commences in a superposition of excited
states, or (iii) undergoes initial oscillations in a superposition
involving all three states of the V model. Quantum interference of
spontaneous emission channels is also integrated into our analysis.
Our findings reveal that the initial state of the atom significantly
influences the potential for achieving 2D spatially dependent spontaneous
emission. In the first case (i), when the atom is initially at its
ground level and without the presence of quantum interference, no
azimuthal dependence is observed in the emission spectrum. However,
the introduction of interference dramatically modifies the scenario,
enabling the imprinting of OAM features from the vortex beam onto
the emission spectrum. This is attributed to quantum interference
closing the level transitions initiated at the ground level, forming
a closed loop. The situation differs for cases (ii) and (iii), where
the spatial features of OAM are transferred to the spontaneous emission
spectrum either with or without quantum interference. Nevertheless,
the presence of interference enhances the results, enabling a diverse
range of spatio-spectral behaviors, including two-dimensional (2D) spectral-peak narrowing,
spectral-peak enhancement, spectral-peak elimination, and spontaneous
emission reduction or quenching in the spatial azimuthal plane. This
research provides valuable insights into azimuthal control of spontaneous
emission through OAM manipulation in optical vortex interactions with
multi-level atomic systems.

Spatial control of spontaneous emission spectra is paramount for precision
engineering and customizing the interaction dynamics between atoms
and photons. In quantum information processing, it enables efficient
quantum gates and enhances computational capabilities through the
precise shaping of emission spectra. In enhanced light-matter interactions,
spatially controlled emission spectra contribute to advanced devices
in quantum optics and cavity quantum electrodynamics, offering tailored
interactions for applications like advanced spectroscopy and imaging.
The importance extends to quantum sensing and metrology, where spatial
control enhances measurement precision and facilitates accurate quantum
state detection. Additionally, in optical communication, creating
specialized optical sources with unique spectral properties enhances
efficiency and security.

It is noteworthy that a closely related scheme has recently been proposed to investigate the generation and detection of optical vortices through spontaneous emission spectra using M-type atoms \citep{Abbas2024}. In our V-type atomic system, the spontaneous emission of light exhibits heightened efficiency compared to an M-type atomic system, primarily due to the distinctive energy level configuration of the involved atoms. Within the V-type atomic system, the configuration involves transitions from a singular lower energy level to two distinct higher energy levels. This unique arrangement facilitates a more efficient and precisely controlled emission of light in contrast to the M-type scheme. The latter entails the superposition of ground state levels, a task that poses challenges in experimental achievement due to technical limitations and inherent complexities. On the other hand, our proposed scheme is specifically designed to achieve spatio-spectral control over spontaneous emission, showcasing a wide range of behaviors. These include two-dimensional spectral-peak narrowing, spectral-peak enhancement, spectral-peak suppression, and spontaneous emission reduction or quenching in the spatial azimuthal plane—effects that were not discussed in Ref. \citep{Abbas2024}.

This paper is organized as follows. In Sec.~\ref{Sec:theory}, we begin our theoretical exploration by elucidating the system configuration and introducing an azimuthally-dependent orbital angular momentum (OAM). In Sec.~\ref{Sec:sds}, we present the solutions detailing the spatial dependence of the spontaneous emission spectrum across various initial states of the atom. Through rigorous examination, we uncover the intricate interplay between the atomic structure and the characteristics of emitted light. Finally, Sec.~\ref{Sec: con} concludes with a summary of our main conclusions and a discussion of their implications providing an insight into the manipulation of emission properties and possible uses of azimuthal control in quantum systems.

\section{Theory}\label{Sec:theory}


\begin{figure}
\includegraphics[width=0.5\columnwidth]{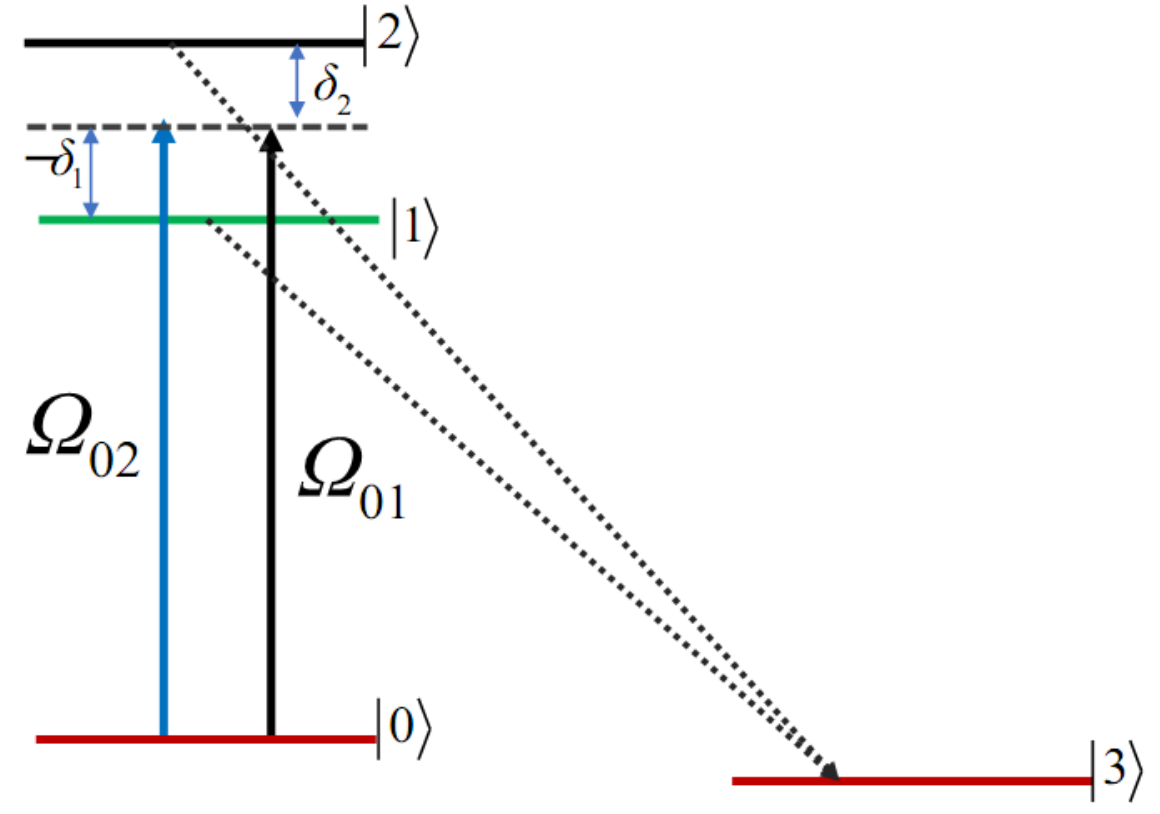}
\caption{Illustration of a four-level atomic system where the ground state
is denoted as $|0\rangle$, coupled to two upper states, $|1\rangle$
and $|2\rangle$, via interaction with two laser fields characterized
by Rabi frequencies $\Omega_{01}$ and $\Omega_{02}$. The upper states
exclusively decay to a common state $|3\rangle$. }
\label{fig:1}
\end{figure}

We consider an atom fixed in position, characterized by a four-level atomic configuration consisting of two lower
levels $|0\rangle$ and $|3\rangle$ and two upper levels $|1\rangle$,
 and $|2\rangle$ as illustrated in Fig.~\ref{fig:1}. In addition one optical
vortex light with Rabi frequency $\varOmega_{01}(r,\phi)$ and a coupling
light with Rabi frequency $\varOmega_{02}$ acts on transitions $|0\rangle\rightarrow|1\rangle$and
$|0\rangle\rightarrow|2\rangle$, respectively. The interaction Hamiltonian within the dipole and Rotating-Wave Approximation, considering two structured waves and coupling fields that drive distinct atomic transitions, is expressed as follows:
\begin{equation}
H_{I}=\hbar\{\Omega_{01}(r,\phi)e^{i\delta_{1}t}|0\rangle\langle1|+\Omega_{02}e^{i\delta_{2}t}|0\rangle\langle2|\}+\sum_{k}\{g_{\textbf{k}_{1}}e^{i(\omega_{13}-\nu_{k})t}a_{k}|1\rangle\langle3|+g_{\textbf{k}_{2}}e^{i(\omega_{23}-\nu_{k})t}a_{k}|2\rangle\langle3|\}+\mathrm{H.c.}\,,\label{eq:1}
\end{equation}
where $a_{k}$ ($a_{k}^{\dagger}$) is the creation (annihilation) operator
of the reservoir modes with frequency $\nu_{k}=ck$ and $g_{\textbf{k}_{1}}$ ($g_{\textbf{k}_{2}}$)
denotes the coupling constant between the vacuum field and the corresponding
atomic transition. The last two terms represent spontaneous emission events, during which photons are emitted in random directions with any polarization. The resonant transition frequencies are
$\omega_{13}$ and $\omega_{23}$. However, the first two terms denote stimulated absorptions of laser photons followed by the subsequent excitation of the atom. The detunings from states $|1\rangle$ and $|2\rangle$ are
denoted by $\delta_{1}$ and $\delta_{2}$ ($\delta_{1(2)}=\omega_{1(2)}-\omega_{0}-\omega$),
where the radiative shifts are omitted. The coupling between the atom and the electromagnetic vacuum field plays a pivotal role in shaping the dynamics of this evolution. Given that the driving field carries optical vortices, the Rabi frequency can be defined as follows:
\begin{equation}
\Omega_{01}(r,\phi)=O_{01}\left(\frac{r}{w}\right)^{|l|}e^{-\frac{r^{2}}{w^{2}}}e^{il\phi}\,,\label{eq:2}
\end{equation}
where $r=\sqrt{x^{2}+y^{2}}$ represents the radial distance from
the beam axis. Here, $\phi$ signifies the azimuthal angle, and $l$
is an integer corresponding to the OAM of light. Additionally, the
parameter $w$ characterizes the beam waist, and $O_{01}$ quantifies
the strength of the vortex and coherent fields. Furthermore, $\ensuremath{\Omega_{01}}(r,\phi)=\ensuremath{\Omega_{10}^{*}}(r,\phi)$ indicating that $\Omega_{10}(r,\phi)=O_{10}\left(\frac{r}{w}\right)^{|l|}e^{-\frac{r^{2}}{w^{2}}}e^{-il\phi}$ and $O_{10}=O_{01}$. In contrast, $\Omega_{02}$ is treated as a real
constant ($\Omega_{02}=\Omega_{20}$), devoid of any spatial dependence. This deliberate choice sets the stage allowing to induce of OAM properties
of light to the spontaneous emission spectrum. In the subsequent section, we explore the spatial characteristics introduced by the azimuthal dependence induced by $\Omega_{01}$. This intentional choice of the state of
light beams serves as a foundation for imparting OAM properties to
the spontaneous emission spectrum. We will explore
the spatial characteristics that emerge from the azimuthal dependence induced by $\Omega_{01}$ on the spontaneous emission of light. 

The state vector of the atom-field system under consideration at time $t$, whose evolution is governed by the Schr\"{o}dinger equation, can be expressed as: 
\begin{equation}
|\varPsi(t)\rangle=\intop drf(r)|r\rangle\Big\{[b_{0}(t)|0\rangle+b_{1}(t)|1\rangle+b_{2}(t)|2\rangle]|\{0\}\rangle+\sum_{k}b_{\textbf{k}}(t)|3\rangle|1_{k}\rangle\Big\}\,,\label{eq:3}
\end{equation}
where the probability amplitude $b_{i}(t)(i=0,1,2)$ represents the
state of atom at time $t$ when there is no spontaneously emitted photon, $b_{\textbf{k}}(t)$
is the probability amplitude that the atom is in level $|3\rangle$
with one photon emitted spontaneously in the $k$th vacuum mode, and
$f(r)$ is the center-of-mass wave function of the atom. In the following calculations, $f(r)$ is presumed to be nearly constant across numerous wavelengths of the vortex light, and it persists unchanged even after interacting with the driving field. The probability amplitude $b_{\textbf{k}}(t\rightarrow\infty)$
can be obtained by solving the Schr\"{o}dinger wave equation with the interaction Hamiltonian Eq.~(\ref{eq:1}) and the atom-field state vector Eq.~(\ref{eq:3}). Making use of the Weisskopf-Wigner theory \citep{Agarwalbook},
the dynamical equations governing the atomic probability amplitudes, with
$\hbar$ set to 1 \citep{Paspalakis98}, are given by
\begin{align}
i\dot{b_{0}}(t)\, & =\Omega_{10}(r,\phi)b_{1}(t)+\Omega_{20}(r,\phi)b_{2}(t),\label{eq:b0}\\
i\dot{b}_{1}(t)\, & =\Omega_{10}(r,\phi)b_{0}(t)+\left(\delta_{1}-i\frac{\Gamma_{1}}{2}\right)b_{1}(t)-ip\frac{\sqrt{\Gamma_{1}\Gamma_{2}}}{2}b_{2}(t),\label{eq:b1}\\
i\dot{b}_{2}(t) & =\Omega_{20}b_{0}(t)-ip\frac{\sqrt{\Gamma_{1}\Gamma_{2}}}{2}b_{1}(t)+
\left(\delta_{2}-i\frac{\Gamma_{2}}{2}\right)b_{2}(t),\label{eq:b2}\\
i\dot{b}_{\textbf{k}}(t) & =\delta_{\textbf{k}}b_{\textbf{k}}(t)-ig_{\textbf{k}_{1}}b_{1}(t)-ig_{\textbf{k}_{2}}b_{2}(t).\label{eq:bk}
\end{align}
Additionally, $\delta_{\textbf{k}}=\omega_{\textbf{k}}-\omega+\omega_{3}-\omega_{0}$,
and $\Gamma_{s}=2\pi|g_{\textbf{k}_{s}}|^{2}D(\omega_{s3})$ for $s=1,2,3,4$,
represent the spontaneous decay rates of states $|1\rangle$ and $|2\rangle$,
respectively. Here, $\textbf{k}$ denotes both the momentum vector and the polarization of the emitted photon. The parameter $D(\omega_{s3})$ signifies the mode density at frequency $\omega_{s3}$. The alignment (of the two dipole moment matrix elements $\vec{ \mu}_{ns}(n=1,2),s=3$) parameter $p$, defined as $p=\vec{\mu}_{13}\cdot\vec{\mu}_{23}/|\vec{\mu}_{13}||\vec{\mu}_{23}|$,
plays a pivotal role in spontaneous emission cancellation \citep{Paspalakis98}.

\subsection{Spontaneous emission spectrum}

The (long-time) spontaneous emission spectrum $S(\delta_{\textbf{q}})$
is expressed as $S(\delta_{\textbf{k}})/S_0=|b_{\textbf{k}}(t\rightarrow\infty)|^{2}$
where $S_0=D(\omega_{s3})$ \citep{Paspalakis98}. We employ the
Laplace transform method to calculate $b_{\textbf{k}}(t\rightarrow\infty)$
\citep{Stephen}. Utilizing Eqs.~(\ref{eq:b1})--(\ref{eq:bk}) and
the final value theorem, one derives the expression
\begin{equation}
b_{\textbf{k}}(t\rightarrow\infty)=-\frac{g_{\textbf{k}_{1}}M(\delta_{\textbf{q}})+g_{\textbf{k}_{2}}N(\delta_{\textbf{q}})}{Z(\delta_{\textbf{k}})},\label{eq:emitted photon}
\end{equation}
where the coefficients $M(\delta_{\textbf{q}})$, $N(\delta_{\textbf{q}})$
and $Z(\delta_{\textbf{k}})$ are given by
\begin{align}
M(\delta_{\textbf{k}}) & =ib_{0}(0)\xi_{0}+ib_{1}(0)\xi_{1}-ib_{2}(0)\xi_{2},\label{eq:eqn1}\\
N(\delta_{\textbf{k}}) & =ib_{0}(0)\xi_{3}-ib_{1}(0)\xi_{4}+ib_{2}(0)\xi_{5},\label{eq:eqn2}\\
Z(\delta_{\textbf{k}}) & =\Lambda+ip\frac{\sqrt{\Gamma_{1}\Gamma_{2}}}{2}\left(\Omega_{10}\Omega_{02}+\Omega_{01}\Omega_{20}\right),\label{eq:eqn3}
\end{align}
with $\xi_{0}=\Omega_{10}X_{2}-i\Omega_{20}p\frac{\sqrt{\Gamma_{1}\Gamma_{2}}}{2}$,
$\xi_{1}=\delta_{\textbf{q}}X_{2}-|\Omega_{02}|^{2}$, $\xi_{2}=i\delta_{\textbf{k}}p\frac{\sqrt{\Gamma_{1}\Gamma_{2}}}{2}-\Omega_{10}\Omega_{02}$,
$\xi_{3}=\Omega_{20}X_{1}-i\Omega_{10}p\frac{\sqrt{\Gamma_{1}\Gamma_{2}}}{2}$,
$\xi_{4}=i\delta_{\textbf{k}}p\frac{\sqrt{\Gamma_{1}\Gamma_{2}}}{2}-\Omega_{01}\Omega_{20}$,
$\xi_{5}=\delta_{\textbf{k}}X_{1}-|\Omega_{01}|^{2}$, $X_{1}=\delta_{\textbf{k}}-\delta_{1}+i\frac{\Gamma_{1}}{2}$,
$X_{2}=\delta_{\textbf{k}}-\delta_{2}+i\frac{\Gamma_{2}}{2}$ and
$\Lambda=\delta_{\textbf{k}}\left(X_{1}X_{2}+p^{2}\frac{\Gamma_{1}\Gamma_{2}}{4}\right)-\left(|\Omega_{01}|^{2}X_{2}+|\Omega_{02}|^{2}X_{1}\right)$.
In this case, the spontaneous emission spectrum $S(\delta_{\textbf{k}})/S_0$
is given
by
\begin{align}
\frac{S(\delta_{\textbf{k}})}{S_0} &=|b_{\textbf{k}}(t\rightarrow\infty)|^{2}=
\left|\frac{g_{\textbf{k}_{1}}M(\delta_{\textbf{k}})+g_{\textbf{k}_{2}}N(\delta_{\textbf{k}})}{Z(\delta_{\textbf{k}})}\right|^{2}\nonumber \\
 & =g_{\textbf{k}_{1}}g_{\textbf{k}_{1}}^{*}\frac{M(\delta_{\textbf{q}})M^{*}(\delta_{\textbf{k}})}{Z(\delta_{\textbf{k}})Z^{*}(\delta_{\textbf{k}})}+g_{\textbf{k}_{2}}g_{\textbf{k}_{2}}^{*}\frac{N(\delta_{\textbf{k}})N^{*}(\delta_{\textbf{k}})}{Z(\delta_{\textbf{k}})Z^{*}(\delta_{\textbf{k}})}+g_{\textbf{k}_{1}}g_{\textbf{k}_{2}}^{*}\frac{M(\delta_{\textbf{k}})N^{*}(\delta_{\textbf{k}})}{Z(\delta_{\textbf{k}})Z^{*}(\delta_{\textbf{k}})}+g_{\textbf{k}_{2}}g_{\textbf{k}_{1}}^{*}\frac{N(\delta_{\textbf{k}})M^{*}(\delta_{\textbf{k}})}{Z(\delta_{\textbf{k}})Z^{*}(\delta_{\textbf{k}})},\label{eq:S}
\end{align}
where $g_{\textbf{k}_{1}}g_{\textbf{k}_{1}}^{*}\propto\Gamma_{1}$, $g_{\textbf{k}_{2}}g_{\textbf{k}_{2}}^{*}\propto\Gamma_{2}$,  and $g_{\textbf{k}_{1}}g_{\textbf{k}_{2}}^{*}=g_{\textbf{k}_{2}}g_{\textbf{k}_{1}}^{*}\propto p\sqrt{\Gamma_{1}\Gamma_{2}}$ \citep{Kapale}. In all the following simulations, we will take $g_{\textbf{k}_{1}}=g_{\textbf{k}_{2}}=1$ \citep{Ghafoor2000}.

\section{Spatially dependent spontaneous emission}\label{Sec:sds}

We will discuss the phenomenon of spatially dependent spontaneous emission in this section, with particular attention to the modification of OAM in the composite vortex light across several atom starting states. We will also investigate the evolution of the emission properties to the different atomic quantum states in terms of quantum
interference. The dynamic interaction between atoms and light will be examined by analyzing the spectral spot patterns in the emission spectrum.


Equation (\ref{eq:S}), along with Eqs.~(\ref{eq:eqn1})--(\ref{eq:eqn3}),
highlights the profound sensitivity of spontaneous emission to the
initial state of the atom. Subsequent analysis explores various potential
initial states for the atom. This includes scenarios where the atom
is initially in its ground state $|0\rangle$, the atom starts in
the superposition of excited states $|1\rangle$ and $|2\rangle$,
and the atom initially oscillates in the superposition of states $|0\rangle$,
$|1\rangle$, and $|2\rangle$. The investigation meticulously assesses
the impact of OAM of light on spontaneous emission in each distinctive
scenario. Our analysis comprehensively considers situations both with
and without the quantum interference term $p$. 

\subsection{Initial state is ground state $|0\rangle$}

Upon assuming that the system's initial state is the ground state
$|0\rangle$, characterized by $b_{0}(0)=1$, $b_{1}(0)=b_{2}(0)=0$,
and with the quantum interference term $p$ being null ($p=0$), the
results gleaned from Eqs.~(\ref{eq:emitted photon}) through
(\ref{eq:eqn3}) reduce to:
\begin{align}
M(\delta_{\textbf{k}}) & =ib_{0}(0)\xi_{0}=i\Omega_{10}X_{2},\nonumber \\
N(\delta_{\textbf{k}}) & =ib_{0}(0)\xi_{3}=i\Omega_{20}X_{1},\nonumber \\
Z(\delta_{\textbf{k}}) & =\delta_{\textbf{k}}X_{1}X_{2}-\left(|\Omega_{01}|^{2}X_{2}+|\Omega_{02}|^{2}X_{1}\right).\label{eq:coefs}
\end{align}

In this particular scenario, where $p=0$,
the final two components in Eq.~(\ref{eq:S}) vanish. Consequently,
only the initial two terms contribute to the spontaneous emission
spectrum. Therefore, the expression of the spontaneous emission spectrum
simplifies to:
\begin{equation}
\frac{S(\delta_{\textbf{k}})}{S_0}=\frac{|\Omega_{01}|^{2}X_{2}X_{2}^{*}}{Z(\delta_{\textbf{k}})Z^{*}(\delta_{\textbf{k}})}+\frac{|\Omega_{02}|^{2}X_{1}X_{1}^{*}}{Z(\delta_{\textbf{k}})Z^{*}(\delta_{\textbf{k}})},\label{eq:S1}
\end{equation}
We can find from Eq.~(\ref{eq:S1}) that the spontaneous emission spectrum just depends significantly on the intensity distribution of the applied
lights. In this case, we have only spatial distribution of the spontaneous
emission spectrum without any dependency on the OAM of the optical
beam $\Omega_{01}$. As a result, $S(\delta_{\textbf{k}})$ remains
unaffected by the OAM in the presence of the vortex beam $\Omega_{01}$.
This insensitivity arises because the magnitude squared of the field
is the only term present in the expression for the spontaneous emission spectrum and it does not incorporate the phase factor $e^{\pm il\phi}$. 

In the presence of quantum interference ($p$), however, the scenario undergoes
a significant transformation. Notably, the expressions
of $M(\delta_{\textbf{q}})$ and $N(\delta_{\textbf{k}})$ become:
\begin{align}
M(\delta_{\textbf{k}}) & =ib_{0}(0)\xi_{0}=i\left(\Omega_{10}X_{2}-i\Omega_{20}p\frac{\sqrt{\Gamma_{1}\Gamma_{2}}}{2}\right),\nonumber \\
N(\delta_{\textbf{k}}) & =ib_{0}(0)\xi_{3}=i\left(\Omega_{20}X_{1}-i\Omega_{10}p\frac{\sqrt{\Gamma_{1}\Gamma_{2}}}{2}\right),\label{eq:coefs2}
\end{align}
and the equation for the $Z(\delta_{\textbf{k}})$ remains consistent
with in Eq.~(\ref{eq:eqn3}). All four components in Eq.~(\ref{eq:S})
actively contribute to shaping the spontaneous emission spectrum,
revealing a discernible dependence on the phase factor $e^{il\phi}$
(or $e^{-il\phi}$). 

In this particular scenario, the angular momentum dependence of $S(\delta_{\bf k})$ now arises from the direct term of $\Omega_{01}$, which appears in $M(\delta_{\textbf{k}})M^{*}(\delta_{\textbf{k}})$, $N(\delta_{\textbf{k}})N^{*}(\delta_{\textbf{k}})$ and $Z(\delta_{\textbf{k}})Z^{*}(\delta_{\textbf{k}})$ 
in Eq.~(\ref{eq:S}).
This indicates
that the spontaneously emitted photon acquires OAM characteristics
of the vortex beam, signifying a transfer of the OAM from the laser
beam $\Omega_{01}$ to the spontaneously emitted photon. The imprinting
of vortices onto the spontaneously emitted photon can be detected
by mapping the spontaneous emission spectrum $S(\delta_{\textbf{k}})$,
as we will illustrate in the following numerical results. The sensitivity
of $S(\delta_{\textbf{k}})$ to the OAM of the beam $\Omega_{01}$
arises from the phase-dependent nature of the four-level system. This
sensitivity becomes apparent when the system is initially prepared
with $b_{0}(0)=1,b_{1}(0)=b_{2}(0)=0$ and in the presence of quantum
interference. Consequently, the spectrum $S(\delta_{\textbf{k}})$
becomes dependent on the azimuthal angle $\phi$ of the vortex beam
carrying the OAM. This sensitivity can be harnessed to measure regions
of spatially varying spontaneous emission by examining $S(\delta_{\textbf{k}})$.
Furthermore, it provides a promising avenue for identifying the winding
number of a vortex beam through mapping the spatially dependent spontaneous
emission spectrum.

\begin{figure}
\includegraphics[width=0.5\columnwidth]{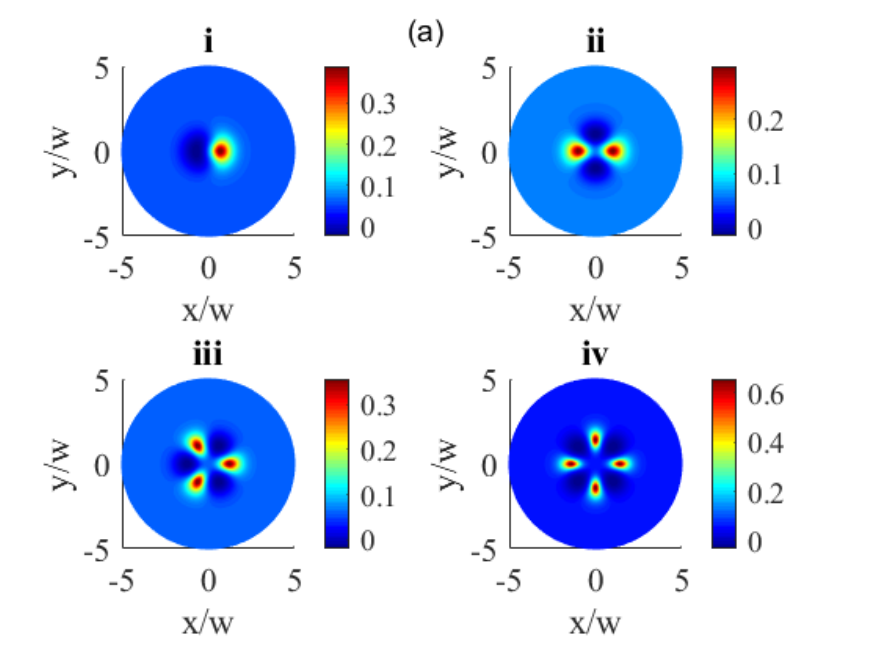}\includegraphics[width=0.5\columnwidth]{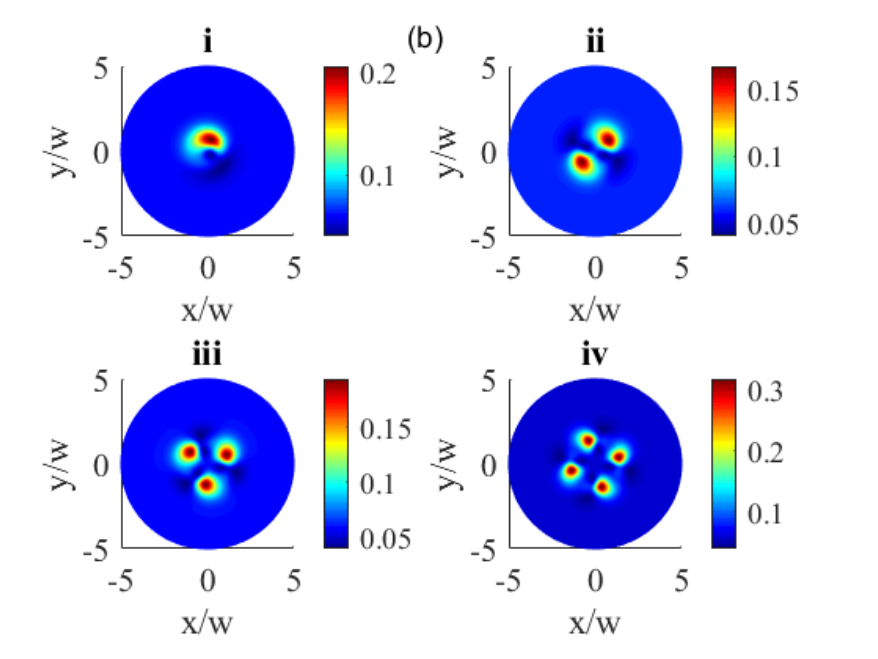}

\caption{Azimuthal modulation of the spontaneous emission $S(\delta_{\textbf{k}})/S_0$
for the first initial condition where $b_{0}(0)=1,b_{1}(0)=b_{2}(0)=0$
and in the presence of quantum interference term $p$ ($p=1$) for
(i) $l=1$, (ii) $l=2$, (iii) $l=3$ and (iv) $l=4$. In panel (a) $\delta_{1}=\delta_{2}=0$,
and hence $\omega_{21}=0$. In panel (b) $\delta_{1}=-\delta_{2}=-\Gamma$
and hence $\omega_{21}=2\Gamma$. The other parameters are $\delta_{\textbf{k}}=0$,
$\Omega_{02}=O_{01}=\Gamma$ and $\Gamma_{1}=\Gamma_{2}=\Gamma$.}
\label{fig:fig2}
\end{figure}

Figure~\ref{fig:fig2} depicts the 2D spontaneous
emission spectrum $S(\delta_{\textbf{k}})$ of an atom initially in
the ground state across various vorticities in the $x-y$ plane. Panel
(a) shows a scenario with degenerate upper states $\omega_{21}=0$,
while panel (b) presents a nondegenerate case $\omega_{21}=2\Gamma$,
where $\omega_{21}=\delta_{2}-\delta_{1}$ represents the energy splitting
of upper states. Throughout all simulations, we maintain $\Gamma_{1}=\Gamma_{2}=\Gamma$
and all parameters are scaled by $\Gamma$. The dark-red structures
in the figure denote positions of spontaneous emission enhancement
or 2D spectral peaks, while the blue areas correspond to regions causing
quenching or reduction in spontaneous emission in the 2D azimuthal
plane. In Fig.\ref{fig:fig2} panel (a) depicting the degenerate case, 
a 2D spectral peak for $l=1$ is embedded in regions of zero spontaneous
emission. The profile of $S(\delta_{\textbf{k}})$
exhibits an $l$-fold symmetry. Increasing the winding
number to larger values augments the number of 2D spectral spots,
while simultaneously modifying their amplitude and width. The width
of the 2D spectral peaks is consistently reduced with increasing $l$, indicating that larger $l$ values result in the narrowing of the spectral
peaks. Notably, the enhancement of 2D spectral peaks is also observed
for $l=4$ (see Fig.~\ref{fig:fig2}(iv), panel (a)). Consequently, determining
an unknown vorticity of a beam $\Omega_{01}$ becomes straightforward
by counting the spectral spot structures in the $S(\delta_{\textbf{k}})$
spectrum. Moreover, the spatially dependent suppression of spontaneous
emission can be established while simultaneously modifying the width
and amplitude of the 2D spectral peaks by tuning the topological charge
$l$. Similar trends are observed for the non-degenerate case, as shown
in Fig.~\ref{fig:fig2}, panel (b). However, the entire spontaneous emission
spectrum is observed to rotate counterclockwise in comparison to Fig.~\ref{fig:fig2},  panel (a).
The spatial control observed in the spontaneous emission spectrum,
leading to regions of spectral peak narrowing or enhancement and quenching
of spontaneous emission, directly stems from the transfer of OAM from
the vortex beam $\Omega_{01}$ to the spontaneous emission.

\subsection{Initial state is a superpostion of excited states $|1\rangle$ and $|2\rangle$}

Considering the initial conditions where the system resides in a superposition
of upper states characterized by $b_{0}(0)=0,b_{1}(0)=b_{2}(0)=\frac{1}{\sqrt{2}}$
, and a null quantum interference term $p=0$, we derive expressions
for $M(\delta_{\textbf{k}})$ and $N(\delta_{\textbf{k}})$ from Eqs.~(\ref{eq:eqn1})--(\ref{eq:eqn3}) 
\begin{align}
M(\delta_{\textbf{k}}) & =ib_{1}(0)\xi_{1}-ib_{2}(0)\xi_{2}=\frac{i}{\sqrt{2}}\left(\xi_{1}+\Omega_{10}\Omega_{02}\right),\nonumber \\
N(\delta_{\textbf{k}}) & =-ib_{1}(0)\xi_{4}+ib_{2}(0)\xi_{5}=\frac{i}{\sqrt{2}}\left(\xi_{5}+\Omega_{01}\Omega_{20}\right),\label{eq:coefss}
\end{align}
and $Z(\delta_{\textbf{k}})$ retains its structure as outlined in
Eq.~(\ref{eq:coefs}). Consequently, one obtains for $S(\delta_{\textbf{k}})$
\begin{equation}
S(\delta_{\textbf{k}})\propto|b_{\textbf{k}}(t\rightarrow\infty)|^{2}=\frac{\xi_{1}\xi_{1}^{*}+\xi_{5}\xi_{5}^{*}+\Omega_{01}\Omega_{20}\left(\xi_{1}+\xi_{5}^{*}\right)+\Omega_{10}\Omega_{02}\left(\xi_{1}^{*}+\xi_{5}\right)+2|\Omega_{01}|^{2}|\Omega_{02}|^{2}}{2Z(\delta_{\textbf{k}})Z^{*}(\delta_{\textbf{k}})}.\label{eq:S1-2}
\end{equation}

\begin{figure}
\includegraphics[width=0.5\columnwidth]{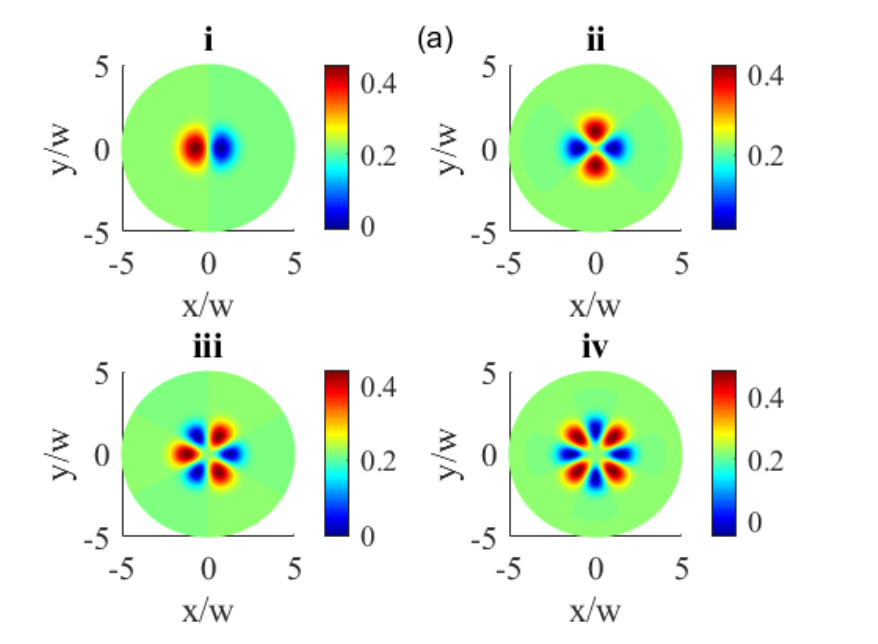}\includegraphics[width=0.5\columnwidth]{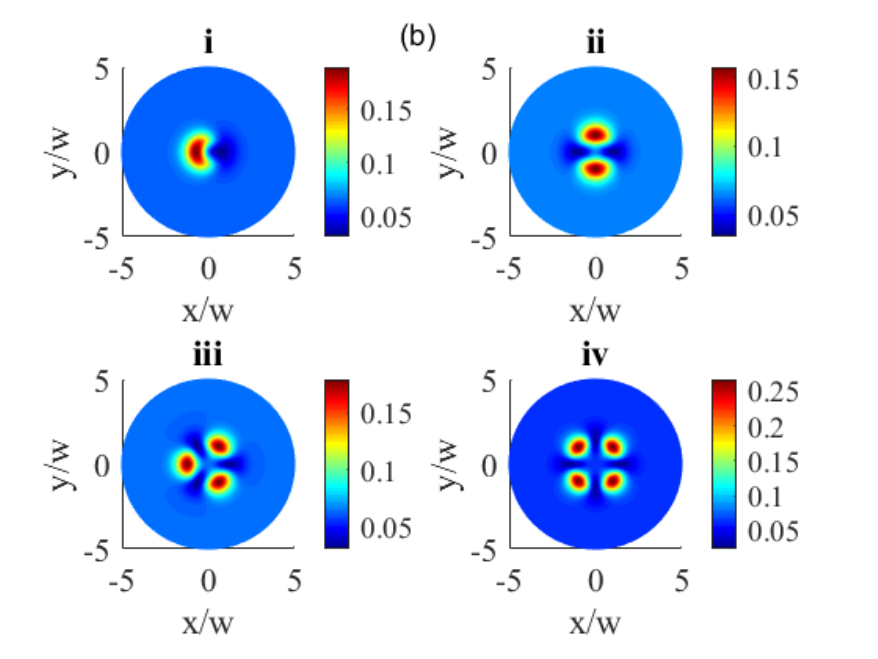}

\caption{Azimuthal modulation of the spontaneous emission $S(\delta_{\textbf{k}})/S_0$
for the second initial condition where $b_{0}(0)=0,b_{1}(0)=b_{2}(0)=\frac{1}{\sqrt{2}}$
and in the absence of quantum interference term $p$ ($p=0$) for
(i) $l=1$, (ii) $l=2$, (iii) $l=3$and (iv) $l=4$. In panel (a) $\delta_{1}=\delta_{2}=0$
and hence $\omega_{21}=0$. In panel (b) $\delta_{1}=-\delta_{2}=-\Gamma$
and hence $\omega_{21}=2\Gamma$. The other parameters are $\delta_{\textbf{k}}=0$,
$\Omega_{02}=O_{01}=\Gamma$ and $\Gamma_{1}=\Gamma_{2}=\Gamma$.}
\label{fig:fig3}
\end{figure}

In this case, even when $p=0$, as indicated by Eq.~(\ref{eq:S1-2}),
a pronounced effect of the optical beam's OAM is evident. This effect
is noticeable owing to the presence of terms $\Omega_{10}\Omega_{02}$
and $\Omega_{10}\Omega_{02}$ in the numerator of Eq. (\ref{eq:S1-2}).
The azimuthally dependent spontaneous emission $S(\delta_{\textbf{k}})$
for this specific case is showcased in Fig.~\ref{fig:fig3} panels (a) and
(b) for degenerate and non-degenerate scenarios, respectively, and
varying vortex numbers. These figures distinctly illustrate the
transference of OAM from the vortex beam to the spontaneous emission
spectrum in the absence of $p$. Upon a closer examination of Figs.~\ref{fig:fig3}(a),
it becomes apparent that both 2D spectral-peak enhancements (highlighted
by red spots) and 2D spectral-peak suppressions (indicated by blue
spots) manifest themselves. These phenomena emerge within a backdrop
of moderate spontaneous emission (highlighted in the green zone).
The number of these enhancements and suppressions increases with the
elevation of the charge $l$, coupled with a narrowing of their respective
widths. The trends observed for the nondegenerate case are similar,
except that in this case spontaneous emission does not completely
diminish to zero.

\begin{figure}
\includegraphics[width=0.5\columnwidth]{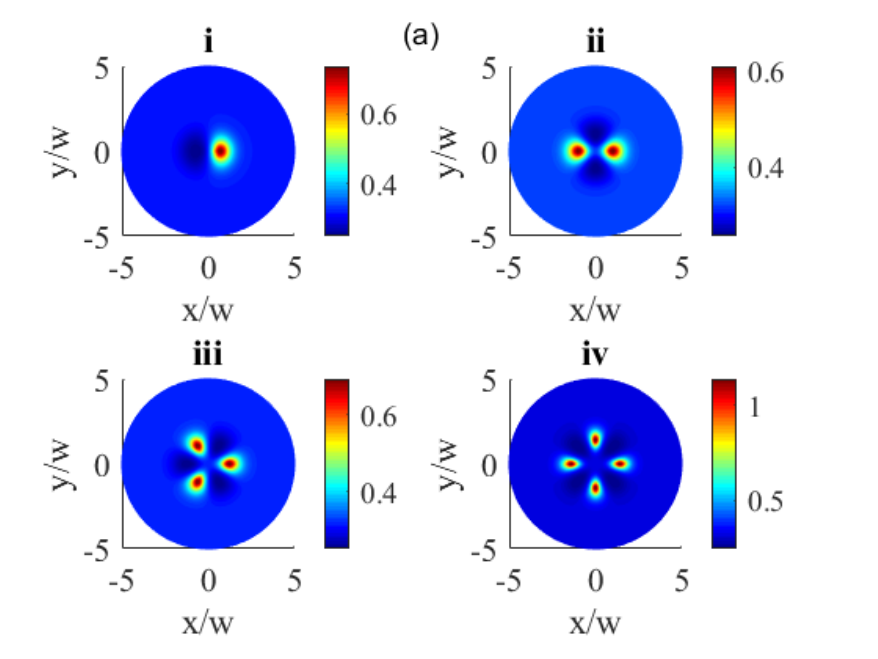}\includegraphics[width=0.5\columnwidth]{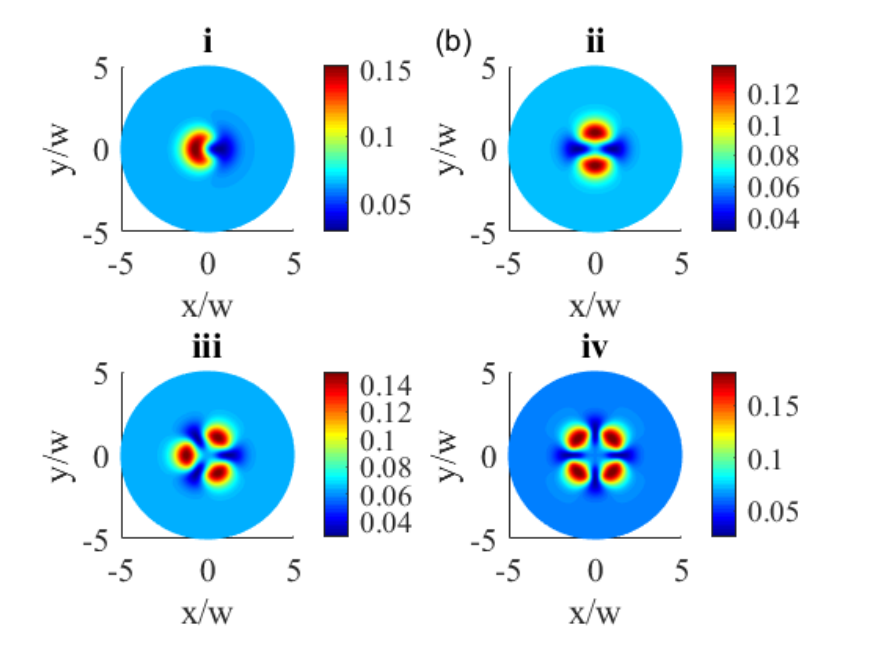}

\caption{Azimuthal modulation of the spontaneous emission $S(\delta_{\textbf{k}})/S_0$
for the last initial condition where $b_{0}(0)=0,b_{1}(0)=b_{2}(0)=\frac{1}{\sqrt{2}}$
and in the presence of quantum interference term $p$ ($p=1$) for
(i) $l=1$, (ii) $l=2$, (iii) $l=3$and (iv) $l=4$. In panel (a) $\delta_{1}=\delta_{2}=0$
and hence $\omega_{21}=0$. In panel (b) $\delta_{1}=-\delta_{2}=-\Gamma$
and hence $\omega_{21}=2\Gamma$. The other parameters are $\delta_{\textbf{k}}=0$,
$\Omega_{02}=O_{01}=\Gamma$ and $\Gamma_{1}=\Gamma_{2}=\Gamma$.}
\label{fig:fig4}
\end{figure}

Exploring a different regime of interest for the initial condition
where the atom resides initially in a superposition of upper states,
we turn our attention to the presence of quantum interference. While
the analytical solutions for $S(\delta_{\textbf{k}})$ are not provided
here due to their extensive nature and lack of inherent informativeness,
they can be directly derived from Eqs.~(\ref{eq:emitted photon})--(\ref{eq:S})
by setting $b_{0}(0)=0,b_{1}(0)=b_{2}(0)=\frac{1}{\sqrt{2}}$ and
$p\neq0$. Opting for maximum quantum interference with $p=1$ (as
similar to Fig.~\ref{fig:fig2}) we examine in Fig.~\ref{fig:fig4}
how interference in this case introduces modifications to the spatially
dependent spontaneous emission spectra. Figure~\ref{fig:fig4} distinctly
reveals the preservation of $l$-fold symmetry in the spectral spots.
Particularly noteworthy is the outcome for the degenerate case $\omega_{21}=0$
and $l=4$, where the 2D spectral peak experiences significant enhancement
alongside a narrowing of the spectral peak (Fig.~\ref{fig:fig4}(iv), panel (a)).
Conversely, in the nondegenerate case $\omega_{21}=2\Gamma$ (Fig.~\ref{fig:fig4},  panel (b)),
the spectral peaks undergo a counterclockwise rotation once again.

\subsection{Initial state is superposition of $|0\rangle$, $|1\rangle$ and $|2\rangle$ states}

\begin{figure}
\includegraphics[width=0.5\columnwidth]{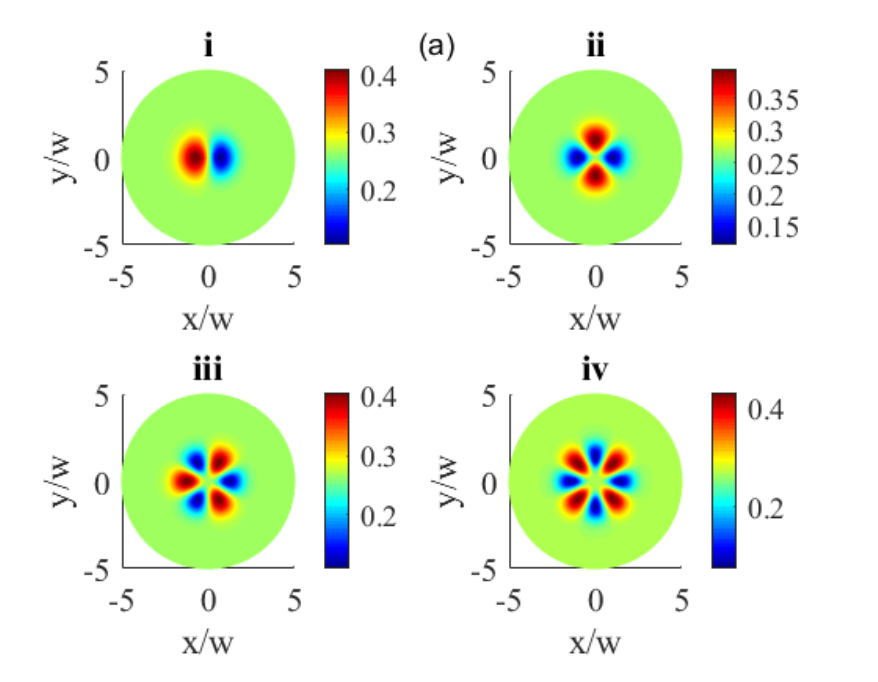}\includegraphics[width=0.5\columnwidth]{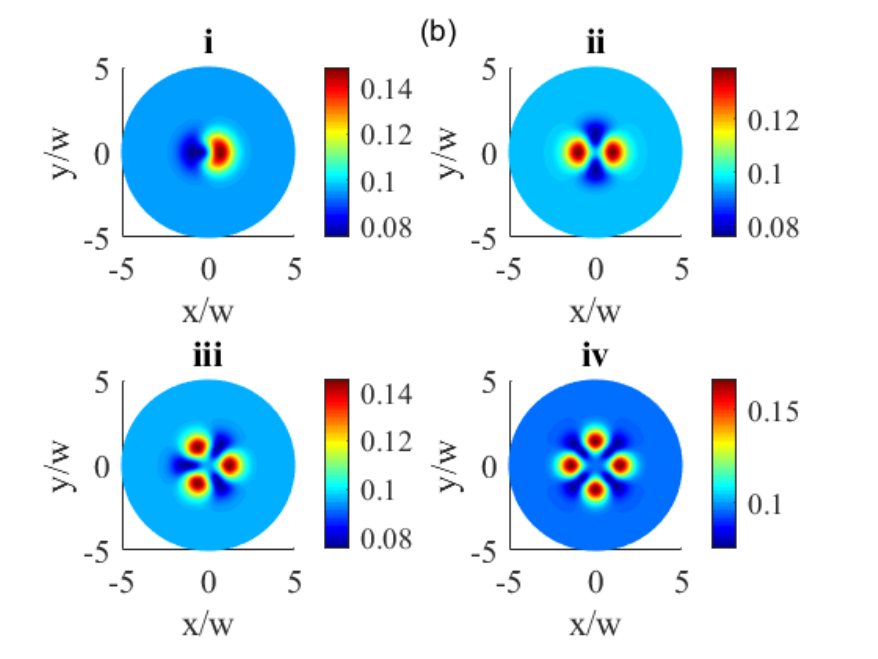}

\caption{Azimuthal modulation of the spontaneous emission $S(\delta_{\textbf{k}})$
for the case where $b_{0}(0)=b_{1}(0)=b_{2}(0)=\frac{1}{\sqrt{3}}$
and in the absence of quantum interference term $p$ ($p=0$) for
(i) $l=1$, (ii) $l=2$, (iii) $l=3$and (iv) $l=4$. In panel (a) $\delta_{1}=\delta_{2}=0$
and hence $\omega_{21}=0$. In panel (b) $\delta_{1}=-\delta_{2}=-\Gamma$
and hence $\omega_{21}=2\Gamma$. The other parameters are $\delta_{\textbf{k}}=0$,
$\Omega_{02}=O_{01}=\Gamma$ and $\Gamma_{1}=\Gamma_{2}=\Gamma$.}
\label{fig:fig5}
\end{figure}

\begin{figure}
\includegraphics[width=0.5\columnwidth]{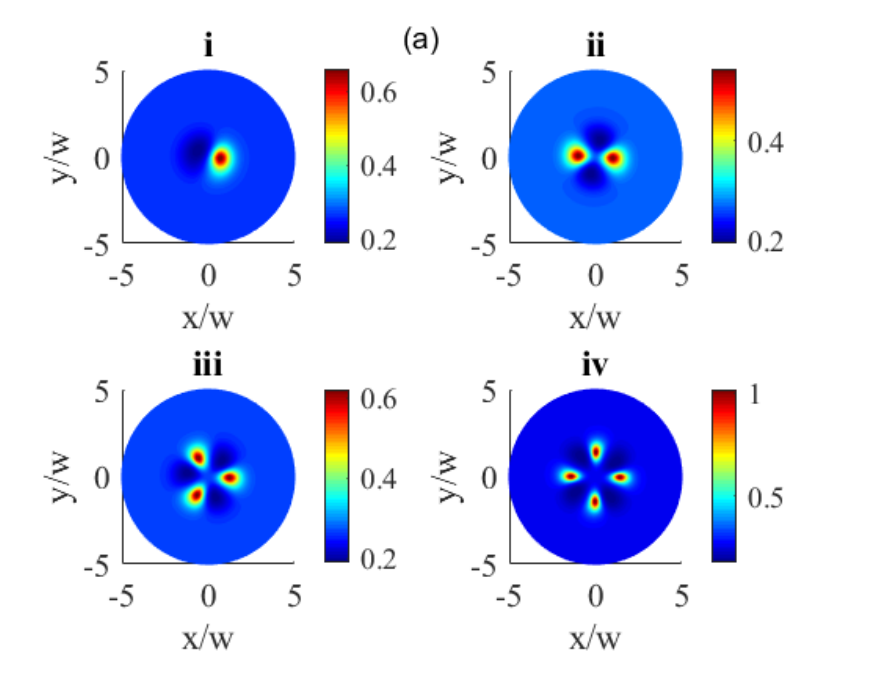}\includegraphics[width=0.5\columnwidth]{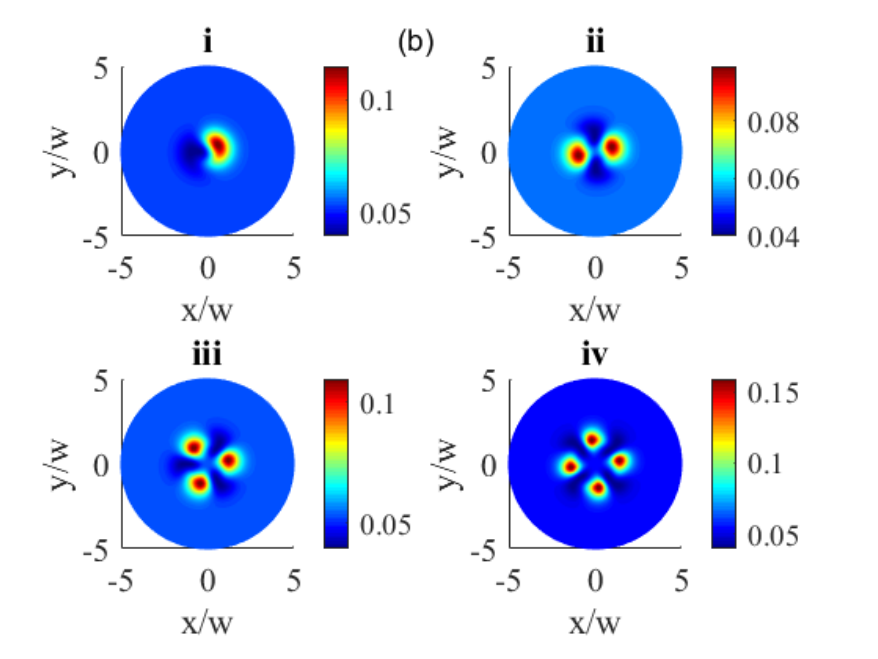}

\caption{Azimuthal modulation of the spontaneous emission $S(\delta_{\textbf{k}})/S_0$
for the case where $b_{0}(0)=b_{1}(0)=b_{2}(0)=\frac{1}{\sqrt{3}}$
and in the presence of quantum interference term $p$ ($p=1$) for
(i) $l=1$, (ii) $l=2$, (iii) $l=3$and (iv) $l=4$. In panel (a) $\delta_{1}=\delta_{2}=0$
and hence $\omega_{21}=0$. In panel (b) $\delta_{1}=-\delta_{2}=-\Gamma$
and hence $\omega_{21}=2\Gamma$. The other parameters are $\delta_{\textbf{k}}=0$,
$\Omega_{02}=O_{01}=\Gamma$ and $\Gamma_{1}=\Gamma_{2}=\Gamma$.}
\label{fig:fig6}
\end{figure}

In a generic scenario, where the atom is initially distributed
among all three states $|0\rangle$, $|1\rangle$ and $|2\rangle$,
the solutions for this case were presented in Eqs.~(\ref{eq:emitted photon})--(\ref{eq:S}).
Figure \ref{fig:fig5} (\ref{fig:fig6}) illustrates the simulation
results for the 2D spatially dependent spontaneous emission in this
context, considering both $p=0$ and $p=1$. Panel (a) in these figures
depicts the degenerate case, while panel (b) showcases the non-degenerate
case.

For this initial superposition state preparation of the atom, the
spontaneous emission exhibits again spatial dependence both in the
absence and presence of quantum interference $p$, imprinting the
OAM characteristics of the vortex beam onto the spectrum of $S(\delta_{\textbf{k}})$.
Particularly interesting phenomena, such as 2D spectral-peak narrowing
and spectral-peak enhancement (see e.g., Fig.~\ref{fig:fig6}(iv) panel (a),
spectral-peak suppression (see e.g., Fig.~\ref{fig:fig5}(a)), and
spontaneous emission reduction or quenching in the azimuthal plane,
can be observed. All these effects are achievable by varying the vorticity
number $l$. Interestingly for $p=0$ ($p=1$), a clockwise (counterclockwise)
rotation of 2D patterns is observed for the non-degenerate case, as
shown in panel (b) Fig.~\ref{fig:fig5} (similarly in Fig.~\ref{fig:fig6}) with
respect to the degenerate case illustrated in panel (a) Fig.~\ref{fig:fig5}
(or in Fig.~\ref{fig:fig6}).\\

\subsection{The case with  Gaussian beams}
Finally, we consider the case where both coupling fields are Gaussian beams with the OAM equal to zero. For this scenario, we analyze the situations depicted in Fig.~7: (i) $p=0$, $b_{0}(0)=1$, $b_{1}(0)=b_{2}(0)=0$; (ii) $p=1$, $b_{0}(0)=1$, $b_{1}(0)=b_{2}(0)=0$; (iii) $p=0$, $b_{0}(0)=0$, $b_{1}(0)=b_{2}(0)=\frac{1}{\sqrt{2}}$; and (iv) $p=1$, $b_{0}(0)=0$, $b_{1}(0)=b_{2}(0)=\frac{1}{\sqrt{2}}$. In panel (a), $\delta_{1}=\delta_{2}=0$, resulting in $\omega_{21}=0$. In panel (b), $\delta_{1}=-\delta_{2}=-\Gamma$, leading to $\omega_{21}=2\Gamma$. The other parameters are $\delta_{\mathbf{k}}=0$, $\Omega_{02}=\Omega_{01}=\Gamma$, and $\Gamma_{1}=\Gamma_{2}=\Gamma$.

As illustrated, when applying Gaussian beams, the spontaneous emission spectrum exhibits no spatial dependency and remains homogeneous in the azimuthal plane. This contrasts with the previous cases involving optical vortices. In the first two scenarios (i) and (ii), where the population initially resides in the ground level ($b_{0}(0)=1$, $b_{1}(0)=b_{2}(0)=0$), the patterns display a ring structure with maximum 2D spectral patterns of spontaneous emission occurring in the rings. Conversely, in scenarios (iii) and (iv), where the initial condition is $b_{0}(0)=0$, $b_{1}(0)=b_{2}(0)=\frac{1}{\sqrt{2}}$, the patterns show a homogeneous distribution of spontaneous emission in the 2D azimuthal space, with spectral-peak suppression at the core of the azimuthal plane.

Furthermore, the results indicate that nonzero detuning does not affect the overall pattern, but only changes the amplitude of the plots, as seen by comparing panels (a) and (b). These findings suggest that the optimal approach to observing spatially dependent spontaneous emission spectra involves using inhomogeneous optical vortex beams, which can transfer their inhomogeneity to the spatio-spectral patterns.

\begin{figure}
\includegraphics[width=0.5\columnwidth]{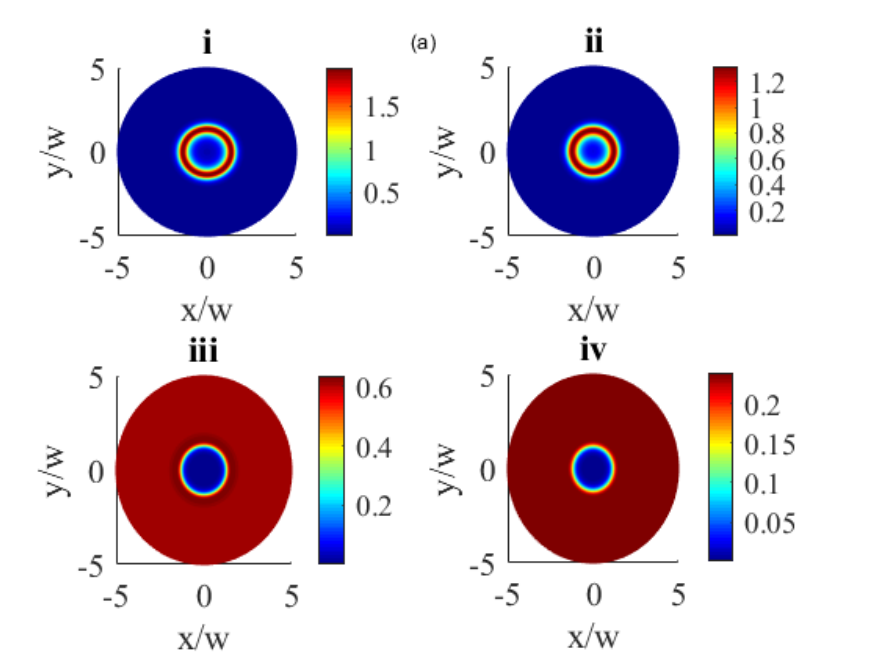}\includegraphics[width=0.5\columnwidth]{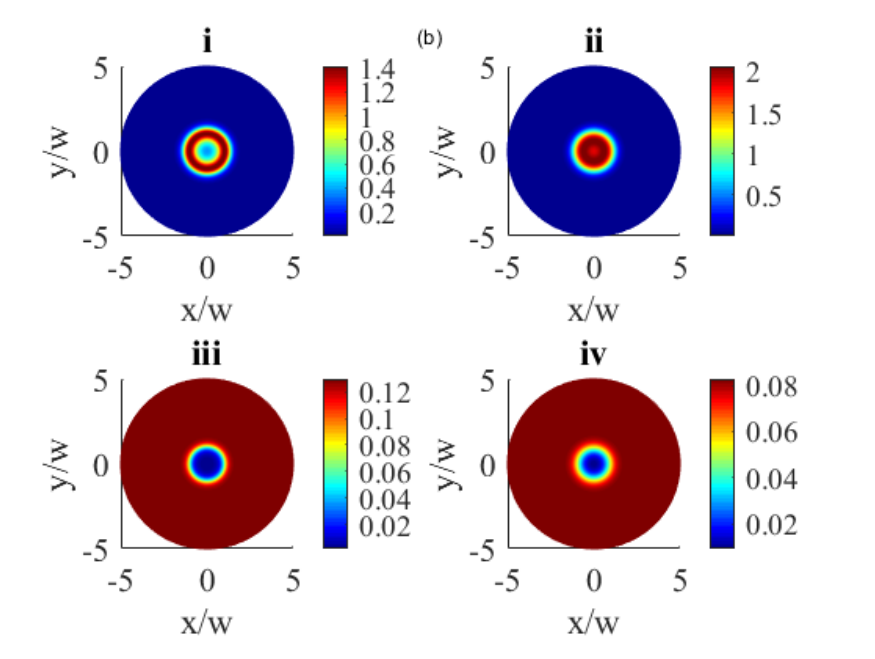}

\caption{Azimuthal modulation of the spontaneous emission $S(\delta_{\textbf{k}})/S_0$
for  the case where both beams are Gaussian beams and (i) $p=0$, $b_{0}(0)=1,b_{1}(0)=b_{2}(0)=0$, (ii) $p=1$, $b_{0}(0)=1,b_{1}(0)=b_{2}(0)=0$, (iii) $p=0$, $b_{0}(0)=0,b_{1}(0)=b_{2}(0)=\frac{1}{\sqrt{2}}$
and (iv) $p=1$, $b_{0}(0)=0,b_{1}(0)=b_{2}(0)=\frac{1}{\sqrt{2}}$. In panel (a) $\delta_{1}=\delta_{2}=0$
and hence $\omega_{21}=0$. In panel (b) $\delta_{1}=-\delta_{2}=-\Gamma$
and hence $\omega_{21}=2\Gamma$. The other parameters are $\delta_{\textbf{k}}=0$,
$\Omega_{02}=O_{01}=\Gamma$ and $\Gamma_{1}=\Gamma_{2}=\Gamma$.}
\label{fig:fig7}
\end{figure}

\subsection{Further discussions and limitations of the model}
In this section, we explore the nature of the observed patterns and the mechanisms behind them. We specifically focus on cases of spatially dependent spontaneous emission illustrated in Figs. 2-6, contrasting with spatially independent outcomes observed with Gaussian beams (Fig. 7).

The modifications and rotations observed in panels (a) and (b) of these figures primarily arise from the introduction of non-zero detunings in panel (b) for the nondegenerate cases. These alterations can be interpreted through the analytical solutions of the spontaneous emission spectrum, as described in Eq. (12) for each specific scenario.

Let us assume, for instance, the case in Fig. 2(a) where $\delta_{1}=0$, $\delta_{2}=0$, $p=1$, and initial conditions $b_{0}(0)=1$, $b_{1}(0)=b_{2}(0)=0$. In this case, Eq. (12) simplifies to:
\begin{equation}
\frac{S(\delta_{k}=0)}{S_{0}} = \frac{1}{2} \frac{|\Omega_{01}|^{2} + |\Omega_{02}|^{2} - 2\Omega\Omega_{02} \cos(l\phi)}{|\Omega_{01}|^{2} + |\Omega_{02}|^{2} + 2\Omega\Omega_{02} \cos(l\phi)|^{2}},
\end{equation}
where we have assumed $\Omega_{01} = \Omega e^{il\phi}$, $\Omega_{10} = \Omega e^{-il\phi}$, $\Omega = \Omega_{01}\left(\frac{r}{w}\right)^{|l|}e^{-\frac{r^{2}}{w^{2}}}$, and $\Omega_{02}$ is a real constant ($\Omega_{02}=\Omega_{20}$). The term $\cos(l\phi)$ introduces $l$-fold symmetry, resulting in a sinusoidal modulation of spontaneous emission patterns in Fig. 2(a), indicating that increasing the OAM number $l$ enhances the number of spectral spots.

Analytical solutions of Eq. (12) for the spontaneous emission spectrum for the nondegenerate case with non-zero detuning provide further insights into further modifications of patterns. While the detailed expressions are extensive in such cases, a general examination of parameters $X_{1}$ and $X_{2}$, defined after Eq. (11), elucidates the rotational nature of these changes. Non-zero detunings for the nondegenerate case introduce real values into these parameters, leading to additional terms in the final spontaneous emission spectrum compared to the degenerate scenario (Eq. (18)). These additional terms, such as $\Omega_{10} X_{2}$ via $\xi_{0}$, introduce phase factors  (e.g., $e^{-il\phi}$), thereby accounting for the observed rotational characteristics of the spectrum. Conversely, terms involving the multiplication of the nonvortex beam $\Omega_{20}$ by detuning (e.g., via term $\xi_{3}$) modify spectral peak amplitudes, resulting in either enhancement or reduction of spontaneous emission across the 2D azimuthal plane.

Notably, certain patterns exhibit striking similarities under varied initial conditions. Specifically, in the presence of quantum interference ($p=1$) and under resonant conditions of the coupling lights ($\delta_{i}=0$ for $i=1,2$), the spatial patterns of the spontaneous emission spectrum display significant resemblance across different values of the OAM and initial conditions (see Fig.~2(a), Fig.~4(a), and Fig.~6(a)). The primary distinction lies in the amplitude of the spontaneous emission spectrum in each case, reflecting diverse population distributions in the initial conditions.

Furthermore, scenarios devoid of quantum interference ($p=0$), where initial states involve superpositions of upper excited states (Fig.~3(a) and Fig.~5(a)), reveal similar spatio-spectral patterns under resonant conditions of the coupling lights. This similarity arises due to the influence of the initial population in the upper excited states on spatio-spectral patterns, common to both scenarios (Fig.~3(a) and Fig.~5(a)). It is important to note that when the initial state is the ground state and $p=0$, the spontaneous emission spectrum lacks OAM dependence, resulting in the absence of spatio-spectral patterns.

Conversely, under off-resonant conditions of the coupling lights ($\delta_{1}=-\Gamma$, $\delta_{2}=\Gamma$), spatio-spectral behaviors of the spontaneous emission spectrum notably differ between scenarios with ($p=1$) and without ($p=0$) quantum interference. As discussed earlier, non-zero detuning introduces additional terms in the final spontaneous emission spectrum, such as the modulation of vortex beams $\Omega_{10}$ ($\Omega_{01}$) or $\Omega_{20}$ by detuning, thereby rotating or modulating the amplitude of the 2D spectra. This variability stems from how initial conditions of the quantum system distribute populations among different states.

In our proposed scheme we have not considered relaxation from excited states $|1\rangle$ and $|2\rangle$ to the ground state $|0\rangle$ which significantly impacts our proposed scheme in multiple ways. Incorporating decay channels from the excited states to the ground state alters population dynamics by enhancing $|0\rangle$ population, thereby depleting the population available for coherent processes in the excited states, potentially reducing scheme efficiency. Moreover, spontaneous emission from these states to $|0\rangle$ introduces decoherence, diminishing coherence among $|0\rangle$, $|1\rangle$, and $|2\rangle$, which may broaden spectral lines and diminish interference effects in the emission spectrum. Additional spectral lines corresponding to $|1\rangle \rightarrow |0\rangle$ and $|2\rangle \rightarrow |0\rangle$ further complicate spectrum interpretation related to decay to common state $|3\rangle$. Thus, we initially overlooked relaxation to $|0\rangle$ to focus on spontaneous emission to $|3\rangle$. Mitigating the effects of these relaxation processes may necessitate techniques like optical pumping for population control or methods to suppress decoherence, such as dynamical decoupling.

 In our calculation of the spontaneous emission spectrum, we did not consider inhomogeneous broadening, including Doppler broadening. Doppler broadening results from the thermal motion of atoms, causing a spread in resonance frequencies and broadening the spectral lines. This effect reduces the clarity and distinctiveness of spectral features, complicating the identification of individual spatio-spectral peaks. Consequently, the interference effects that are crucial to our scheme may become less prominent in the spontaneous emission spectrum. Moreover, Doppler broadening affects the population dynamics of the system. The broadened linewidths imply that not all atoms in the ensemble resonate with the laser fields, leading to varying coupling strengths. This discrepancy can create an uneven distribution of populations across states, disrupting the balance necessary for the optimal operation of the scheme. Since these effects diminish the visibility of our results, we have excluded them from our calculations. To enhance precision and control in experimental setups, techniques such as magneto-optical trapping (MOT) can be employed to mitigate Doppler broadening by reducing the thermal motion of atoms, as noted in \cite{Uhlenberg2000MagnetoopticalTO}. This approach significantly improves the practical feasibility of our scheme. Additionally, other experimental methods, such as using narrow-linewidth lasers and velocity-selective pumping techniques, can also mitigate the effects of inhomogeneous broadening.



\subsection{Experimental realization of the proposed model}

To ensure precise phase control, we recommend employing an optical phase-locked loop approach \citep{Friederich2010PhaselockingOT}. Additionally, the effectiveness of our approach is influenced by parameters related to quantum interference terms. Regarding experimental implementation of the proposed setup, we note that quantum interference effects in spontaneous emission have been experimentally observed using bare states of specific   atoms \citep{Norris,Heeg}, molecules \citep{Xia},  and semiconductor quantum dots \citep{Dutt2005}. 

For the proposal of the present article, one may achieve this experimentally by building on the successful work of Xia et al.\ \citep{Xia} with sodium dimers. For our purposes, an additional laser would be introduced to facilitate coupling via a four-photon transition.
Another suitable platform for this experimental realization is the dressed \(^{85}\text{Rb}\), see for example Refs. \citep{Wang2008,Wang2009}. Within this proposal, we can utilize the states \(|5\,P_{1/2}, F=3\rangle\) and \(|5\,D_{3/2}, F=4\rangle\) when dressed by a laser field as states \(|1\rangle\) and \(|2\rangle\). Additionally, \(|5\, S_{1/2}, F=3\rangle\) and \(|5\, S_{1/2}, F=2\rangle\) of \(^{85}\text{Rb}\) can be used as \(|0\rangle\) and \(|3\rangle\), respectively. Here, \(|0\rangle\) represents the ground state, and \(|3\rangle\) corresponds to the intermediate state where the spontaneously emitted structured light spectrum is examined. Then, the dressed atomic system interacts with two additional laser fields for creating the phenomena of this work.  Moreover, another approach is to use a double-dressed InGaAs quantum dot, similar to the work of Ref. \citep{He2008}, for creating the necessary quantum interference.  

Furthermore, another approach for creating the necessary quantum interference in spontaneous emission is to use the idea of simulating quantum interference effects in spontaneous emission using the anisotropic Purcell effect occurring when placing a quantum system with orthogonal electric dipole matrix elements in an anisotropic photonic vacuum \citep{Agarwal2000,YangPRL08,Yannopapas,Jha2015,Hughes,Karanikolas}.


\section{Conclusion} \label{Sec: con}

To summarize, our proposed method for achieving spatio-spectral
control over spontaneous emission in a four-level atom-light coupling
system interacting with optical vortices bearing orbital angular momentum
(OAM), has revealed a panorama of results. Delving into diverse scenarios
and considering the initial state of the atom, this paper has illuminated
an array of spatial and spectral behaviors. These encompass intriguing
features such as two-dimensional (2D) spectral-peak narrowing, enhancement,
suppression, and spontaneous emission reduction or quenching spanning
the spatial azimuthal plane. The inhomogeneous atom-light interaction,
influenced by the optical vortex, has intricately structured these
effects. The transfer of OAM characteristics onto the emission spectrum
not only underscores the potential for azimuthal control of spontaneous
emission but also opens avenues for innovative applications in high
dimensional quantum information processing, enhanced communication
protocols, and novel sensing technologies. This study enriches comprehension
of the dynamics of spatio-spectral phenomena and charts a course for
applications leveraging precise OAM-based manipulation
of spontaneous emission in multi-level atomic systems.

\begin{acknowledgments}
R. A. received partial funding from the Iran National Science Foundation (INSF) under project No. 4026871.
\end{acknowledgments}

\section*{}


\begin{thebibliography}{78}%
\makeatletter
\providecommand \@ifxundefined [1]{%
 \@ifx{#1\undefined}
}%
\providecommand \@ifnum [1]{%
 \ifnum #1\expandafter \@firstoftwo
 \else \expandafter \@secondoftwo
 \fi
}%
\providecommand \@ifx [1]{%
 \ifx #1\expandafter \@firstoftwo
 \else \expandafter \@secondoftwo
 \fi
}%
\providecommand \natexlab [1]{#1}%
\providecommand \enquote  [1]{``#1''}%
\providecommand \bibnamefont  [1]{#1}%
\providecommand \bibfnamefont [1]{#1}%
\providecommand \citenamefont [1]{#1}%
\providecommand \href@noop [0]{\@secondoftwo}%
\providecommand \href [0]{\begingroup \@sanitize@url \@href}%
\providecommand \@href[1]{\@@startlink{#1}\@@href}%
\providecommand \@@href[1]{\endgroup#1\@@endlink}%
\providecommand \@sanitize@url [0]{\catcode `\\12\catcode `\$12\catcode
  `\&12\catcode `\#12\catcode `\^12\catcode `\_12\catcode `\%12\relax}%
\providecommand \@@startlink[1]{}%
\providecommand \@@endlink[0]{}%
\providecommand \url  [0]{\begingroup\@sanitize@url \@url }%
\providecommand \@url [1]{\endgroup\@href {#1}{\urlprefix }}%
\providecommand \urlprefix  [0]{URL }%
\providecommand \Eprint [0]{\href }%
\providecommand \doibase [0]{https://doi.org/}%
\providecommand \selectlanguage [0]{\@gobble}%
\providecommand \bibinfo  [0]{\@secondoftwo}%
\providecommand \bibfield  [0]{\@secondoftwo}%
\providecommand \translation [1]{[#1]}%
\providecommand \BibitemOpen [0]{}%
\providecommand \bibitemStop [0]{}%
\providecommand \bibitemNoStop [0]{.\EOS\space}%
\providecommand \EOS [0]{\spacefactor3000\relax}%
\providecommand \BibitemShut  [1]{\csname bibitem#1\endcsname}%
\let\auto@bib@innerbib\@empty
\bibitem [{\citenamefont {Gerry}\ and\ \citenamefont
  {Knight}(2023)}]{gerry2023introductory}%
  \BibitemOpen
  \bibfield  {author} {\bibinfo {author} {\bibfnamefont {C.~C.}\ \bibnamefont
  {Gerry}}\ and\ \bibinfo {author} {\bibfnamefont {P.~L.}\ \bibnamefont
  {Knight}},\ }\href@noop {} {\emph {\bibinfo {title} {Introductory quantum
  optics}}}\ (\bibinfo  {publisher} {Cambridge university press},\ \bibinfo
  {year} {2023})\BibitemShut {NoStop}%
\bibitem [{\citenamefont {Meystre}\ and\ \citenamefont
  {Scully}(2021)}]{meystre2021quantum}%
  \BibitemOpen
  \bibfield  {author} {\bibinfo {author} {\bibfnamefont {P.}~\bibnamefont
  {Meystre}}\ and\ \bibinfo {author} {\bibfnamefont {M.~O.}\ \bibnamefont
  {Scully}},\ }\href@noop {} {\emph {\bibinfo {title} {Quantum optics}}}\
  (\bibinfo  {publisher} {Springer},\ \bibinfo {year} {2021})\BibitemShut
  {NoStop}%
\bibitem [{\citenamefont {Zhou}\ and\ \citenamefont {Swain}(1996)}]{ZhouSE96}%
  \BibitemOpen
  \bibfield  {author} {\bibinfo {author} {\bibfnamefont {P.}~\bibnamefont
  {Zhou}}\ and\ \bibinfo {author} {\bibfnamefont {S.}~\bibnamefont {Swain}},\
  }\bibfield  {title} {\bibinfo {title} {Ultranarrow spectral lines via quantum
  interference},\ }\href {https://doi.org/10.1103/PhysRevLett.77.3995}
  {\bibfield  {journal} {\bibinfo  {journal} {Phys. Rev. Lett.}\ }\textbf
  {\bibinfo {volume} {77}},\ \bibinfo {pages} {3995} (\bibinfo {year}
  {1996})}\BibitemShut {NoStop}%
\bibitem [{\citenamefont {John}\ and\ \citenamefont {Quang}(1997)}]{John}%
  \BibitemOpen
  \bibfield  {author} {\bibinfo {author} {\bibfnamefont {S.}~\bibnamefont
  {John}}\ and\ \bibinfo {author} {\bibfnamefont {T.}~\bibnamefont {Quang}},\
  }\bibfield  {title} {\bibinfo {title} {Collective switching and inversion
  without fluctuation of two-level atoms in confined photonic systems},\ }\href
  {https://doi.org/10.1103/PhysRevLett.78.1888} {\bibfield  {journal} {\bibinfo
   {journal} {Phys. Rev. Lett.}\ }\textbf {\bibinfo {volume} {78}},\ \bibinfo
  {pages} {1888} (\bibinfo {year} {1997})}\BibitemShut {NoStop}%
\bibitem [{\citenamefont {Zhou}\ and\ \citenamefont {Swain}(1997)}]{Swain97}%
  \BibitemOpen
  \bibfield  {author} {\bibinfo {author} {\bibfnamefont {P.}~\bibnamefont
  {Zhou}}\ and\ \bibinfo {author} {\bibfnamefont {S.}~\bibnamefont {Swain}},\
  }\bibfield  {title} {\bibinfo {title} {Absorption spectrum and gain without
  inversion of a driven two-level atom with arbitrary probe intensity in a
  squeezed vacuum},\ }\href {https://doi.org/10.1103/PhysRevA.55.772}
  {\bibfield  {journal} {\bibinfo  {journal} {Phys. Rev. A}\ }\textbf {\bibinfo
  {volume} {55}},\ \bibinfo {pages} {772} (\bibinfo {year} {1997})}\BibitemShut
  {NoStop}%
\bibitem [{\citenamefont {Paspalakis}\ and\ \citenamefont
  {Knight}(1998)}]{Paspalakis98}%
  \BibitemOpen
  \bibfield  {author} {\bibinfo {author} {\bibfnamefont {E.}~\bibnamefont
  {Paspalakis}}\ and\ \bibinfo {author} {\bibfnamefont {P.~L.}\ \bibnamefont
  {Knight}},\ }\bibfield  {title} {\bibinfo {title} {Phase control of
  spontaneous emission},\ }\href {https://doi.org/10.1103/PhysRevLett.81.293}
  {\bibfield  {journal} {\bibinfo  {journal} {Phys. Rev. Lett.}\ }\textbf
  {\bibinfo {volume} {81}},\ \bibinfo {pages} {293} (\bibinfo {year}
  {1998})}\BibitemShut {NoStop}%
\bibitem [{\citenamefont {Ghafoor}\ \emph {et~al.}(2000)\citenamefont
  {Ghafoor}, \citenamefont {Zhu},\ and\ \citenamefont {Zubairy}}]{Ghafoor2000}%
  \BibitemOpen
  \bibfield  {author} {\bibinfo {author} {\bibfnamefont {F.}~\bibnamefont
  {Ghafoor}}, \bibinfo {author} {\bibfnamefont {S.-Y.}\ \bibnamefont {Zhu}},\
  and\ \bibinfo {author} {\bibfnamefont {M.~S.}\ \bibnamefont {Zubairy}},\
  }\bibfield  {title} {\bibinfo {title} {Amplitude and phase control of
  spontaneous emission},\ }\href {https://doi.org/10.1103/PhysRevA.62.013811}
  {\bibfield  {journal} {\bibinfo  {journal} {Phys. Rev. A}\ }\textbf {\bibinfo
  {volume} {62}},\ \bibinfo {pages} {013811} (\bibinfo {year}
  {2000})}\BibitemShut {NoStop}%
\bibitem [{\citenamefont {Ghafoor}\ \emph {et~al.}(2002)\citenamefont
  {Ghafoor}, \citenamefont {Qamar},\ and\ \citenamefont
  {Zubairy}}]{Ghafoor2002}%
  \BibitemOpen
  \bibfield  {author} {\bibinfo {author} {\bibfnamefont {F.}~\bibnamefont
  {Ghafoor}}, \bibinfo {author} {\bibfnamefont {S.}~\bibnamefont {Qamar}},\
  and\ \bibinfo {author} {\bibfnamefont {M.~S.}\ \bibnamefont {Zubairy}},\
  }\bibfield  {title} {\bibinfo {title} {Atom localization via phase and
  amplitude control of the driving field},\ }\href
  {https://doi.org/10.1103/PhysRevA.65.043819} {\bibfield  {journal} {\bibinfo
  {journal} {Phys. Rev. A}\ }\textbf {\bibinfo {volume} {65}},\ \bibinfo
  {pages} {043819} (\bibinfo {year} {2002})}\BibitemShut {NoStop}%
\bibitem [{\citenamefont {Ant\'on}\ \emph {et~al.}(2005)\citenamefont
  {Ant\'on}, \citenamefont {Calder\'on},\ and\ \citenamefont
  {Carre\~no}}]{anton2005}%
  \BibitemOpen
  \bibfield  {author} {\bibinfo {author} {\bibfnamefont {M.~A.}\ \bibnamefont
  {Ant\'on}}, \bibinfo {author} {\bibfnamefont {O.~G.}\ \bibnamefont
  {Calder\'on}},\ and\ \bibinfo {author} {\bibfnamefont {F.}~\bibnamefont
  {Carre\~no}},\ }\bibfield  {title} {\bibinfo {title} {Spontaneously generated
  coherence effects in a laser-driven four-level atomic system},\ }\href
  {https://doi.org/10.1103/PhysRevA.72.023809} {\bibfield  {journal} {\bibinfo
  {journal} {Phys. Rev. A}\ }\textbf {\bibinfo {volume} {72}},\ \bibinfo
  {pages} {023809} (\bibinfo {year} {2005})}\BibitemShut {NoStop}%
\bibitem [{\citenamefont {Li}\ \emph {et~al.}(2005)\citenamefont {Li},
  \citenamefont {Gao}, \citenamefont {Wu},\ and\ \citenamefont {Wang}}]{AiJun}%
  \BibitemOpen
  \bibfield  {author} {\bibinfo {author} {\bibfnamefont {A.-J.}\ \bibnamefont
  {Li}}, \bibinfo {author} {\bibfnamefont {J.-Y.}\ \bibnamefont {Gao}},
  \bibinfo {author} {\bibfnamefont {J.-H.}\ \bibnamefont {Wu}},\ and\ \bibinfo
  {author} {\bibfnamefont {L.}~\bibnamefont {Wang}},\ }\bibfield  {title}
  {\bibinfo {title} {Simulating spontaneously generated coherence in a
  four-level atomic system},\ }\href@noop {} {\bibfield  {journal} {\bibinfo
  {journal} {J. Phys. B: At. Mol. Opt. Phys.}\ }\textbf {\bibinfo {volume}
  {38}},\ \bibinfo {pages} {3815} (\bibinfo {year} {2005})}\BibitemShut
  {NoStop}%
\bibitem [{\citenamefont {Wu}\ \emph {et~al.}(2005)\citenamefont {Wu},
  \citenamefont {Li}, \citenamefont {Ding}, \citenamefont {Zhao},\ and\
  \citenamefont {Gao}}]{Wu2005}%
  \BibitemOpen
  \bibfield  {author} {\bibinfo {author} {\bibfnamefont {J.-H.}\ \bibnamefont
  {Wu}}, \bibinfo {author} {\bibfnamefont {A.-J.}\ \bibnamefont {Li}}, \bibinfo
  {author} {\bibfnamefont {Y.}~\bibnamefont {Ding}}, \bibinfo {author}
  {\bibfnamefont {Y.-C.}\ \bibnamefont {Zhao}},\ and\ \bibinfo {author}
  {\bibfnamefont {J.-Y.}\ \bibnamefont {Gao}},\ }\bibfield  {title} {\bibinfo
  {title} {Control of spontaneous emission from a coherently driven four-level
  atom},\ }\href {https://doi.org/10.1103/PhysRevA.72.023802} {\bibfield
  {journal} {\bibinfo  {journal} {Phys. Rev. A}\ }\textbf {\bibinfo {volume}
  {72}},\ \bibinfo {pages} {023802} (\bibinfo {year} {2005})}\BibitemShut
  {NoStop}%
\bibitem [{\citenamefont {Fountoulakis}\ \emph {et~al.}(2006)\citenamefont
  {Fountoulakis}, \citenamefont {Terzis},\ and\ \citenamefont
  {Paspalakis}}]{Fountoulakis}%
  \BibitemOpen
  \bibfield  {author} {\bibinfo {author} {\bibfnamefont {A.}~\bibnamefont
  {Fountoulakis}}, \bibinfo {author} {\bibfnamefont {A.~F.}\ \bibnamefont
  {Terzis}},\ and\ \bibinfo {author} {\bibfnamefont {E.}~\bibnamefont
  {Paspalakis}},\ }\bibfield  {title} {\bibinfo {title} {Coherent phenomena due
  to double-dark states in a system with decay interference},\ }\href
  {https://doi.org/10.1103/PhysRevA.73.033811} {\bibfield  {journal} {\bibinfo
  {journal} {Phys. Rev. A}\ }\textbf {\bibinfo {volume} {73}},\ \bibinfo
  {pages} {033811} (\bibinfo {year} {2006})}\BibitemShut {NoStop}%
\bibitem [{\citenamefont {Paspalakis}\ and\ \citenamefont
  {Knight}(2000)}]{PaspalakisJMO}%
  \BibitemOpen
  \bibfield  {author} {\bibinfo {author} {\bibfnamefont {E.}~\bibnamefont
  {Paspalakis}}\ and\ \bibinfo {author} {\bibfnamefont {P.~L.}\ \bibnamefont
  {Knight}},\ }\bibfield  {title} {\bibinfo {title} {Coherent control of
  spontaneous emission in a four-level system},\ }\href@noop {} {\bibfield
  {journal} {\bibinfo  {journal} {Journal of Modern Optics Volume , - Issue 6}\
  }\textbf {\bibinfo {volume} {47}},\ \bibinfo {pages} {1025} (\bibinfo {year}
  {2000})}\BibitemShut {NoStop}%
\bibitem [{\citenamefont {Jia-Hua}\ \emph {et~al.}(2006)\citenamefont
  {Jia-Hua}, \citenamefont {Ji-Bing}, \citenamefont {Ai-Xi},\ and\
  \citenamefont {Chun-Chao}}]{JiaHua}%
  \BibitemOpen
  \bibfield  {author} {\bibinfo {author} {\bibfnamefont {L.}~\bibnamefont
  {Jia-Hua}}, \bibinfo {author} {\bibfnamefont {L.}~\bibnamefont {Ji-Bing}},
  \bibinfo {author} {\bibfnamefont {C.}~\bibnamefont {Ai-Xi}},\ and\ \bibinfo
  {author} {\bibfnamefont {Q.}~\bibnamefont {Chun-Chao}},\ }\bibfield  {title}
  {\bibinfo {title} {Spontaneous emission spectra and simulating multiple
  spontaneous generation coherence in a five-level atomic medium},\ }\href
  {https://doi.org/10.1103/PhysRevA.74.033816} {\bibfield  {journal} {\bibinfo
  {journal} {Phys. Rev. A}\ }\textbf {\bibinfo {volume} {74}},\ \bibinfo
  {pages} {033816} (\bibinfo {year} {2006})}\BibitemShut {NoStop}%
\bibitem [{\citenamefont {Arun}(2008)}]{Arun}%
  \BibitemOpen
  \bibfield  {author} {\bibinfo {author} {\bibfnamefont {R.}~\bibnamefont
  {Arun}},\ }\bibfield  {title} {\bibinfo {title} {Interference-induced
  splitting of resonances in spontaneous emission},\ }\href
  {https://doi.org/10.1103/PhysRevA.77.033820} {\bibfield  {journal} {\bibinfo
  {journal} {Phys. Rev. A}\ }\textbf {\bibinfo {volume} {77}},\ \bibinfo
  {pages} {033820} (\bibinfo {year} {2008})}\BibitemShut {NoStop}%
\bibitem [{\citenamefont {Li}\ \emph {et~al.}(2008)\citenamefont {Li},
  \citenamefont {Song}, \citenamefont {Wei}, \citenamefont {Wang},\ and\
  \citenamefont {Gao}}]{Li08}%
  \BibitemOpen
  \bibfield  {author} {\bibinfo {author} {\bibfnamefont {A.-J.}\ \bibnamefont
  {Li}}, \bibinfo {author} {\bibfnamefont {X.-L.}\ \bibnamefont {Song}},
  \bibinfo {author} {\bibfnamefont {X.-G.}\ \bibnamefont {Wei}}, \bibinfo
  {author} {\bibfnamefont {L.}~\bibnamefont {Wang}},\ and\ \bibinfo {author}
  {\bibfnamefont {J.-Y.}\ \bibnamefont {Gao}},\ }\bibfield  {title} {\bibinfo
  {title} {Effects of spontaneously generated coherence in a microwave-driven
  four-level atomic system},\ }\href
  {https://doi.org/10.1103/PhysRevA.77.053806} {\bibfield  {journal} {\bibinfo
  {journal} {Phys. Rev. A}\ }\textbf {\bibinfo {volume} {77}},\ \bibinfo
  {pages} {053806} (\bibinfo {year} {2008})}\BibitemShut {NoStop}%
\bibitem [{\citenamefont {Wang}\ \emph {et~al.}(2012)\citenamefont {Wang},
  \citenamefont {Tian},\ and\ \citenamefont {Wu}}]{ChunOE}%
  \BibitemOpen
  \bibfield  {author} {\bibinfo {author} {\bibfnamefont {C.~L.}\ \bibnamefont
  {Wang}}, \bibinfo {author} {\bibfnamefont {Z.~H. K. S.~C.}\ \bibnamefont
  {Tian}},\ and\ \bibinfo {author} {\bibfnamefont {J.~H.}\ \bibnamefont {Wu}},\
  }\bibfield  {title} {\bibinfo {title} {Control of spontaneous emission from a
  micro-wave driven atomic system},\ }\href@noop {} {\bibfield  {journal}
  {\bibinfo  {journal} {Optics Express}\ }\textbf {\bibinfo {volume} {20}},\
  \bibinfo {pages} {3509} (\bibinfo {year} {2012})}\BibitemShut {NoStop}%
\bibitem [{\citenamefont {Thanopulos}\ \emph {et~al.}(2022)\citenamefont
  {Thanopulos}, \citenamefont {Karanikolas},\ and\ \citenamefont
  {Paspalakis}}]{Thanopulos}%
  \BibitemOpen
  \bibfield  {author} {\bibinfo {author} {\bibfnamefont {I.}~\bibnamefont
  {Thanopulos}}, \bibinfo {author} {\bibfnamefont {V.}~\bibnamefont
  {Karanikolas}},\ and\ \bibinfo {author} {\bibfnamefont {E.}~\bibnamefont
  {Paspalakis}},\ }\bibfield  {title} {\bibinfo {title} {Spontaneous emission
  of a quantum emitter near a graphene nanodisk under strong light-matter
  coupling},\ }\href {https://doi.org/10.1103/PhysRevA.106.013718} {\bibfield
  {journal} {\bibinfo  {journal} {Phys. Rev. A}\ }\textbf {\bibinfo {volume}
  {106}},\ \bibinfo {pages} {013718} (\bibinfo {year} {2022})}\BibitemShut
  {NoStop}%
\bibitem [{\citenamefont {Thanopulos}\ \emph {et~al.}(2019)\citenamefont
  {Thanopulos}, \citenamefont {Karanikolas}, \citenamefont {Iliopoulos},\ and\
  \citenamefont {Paspalakis}}]{Thanopulos2019}%
  \BibitemOpen
  \bibfield  {author} {\bibinfo {author} {\bibfnamefont {I.}~\bibnamefont
  {Thanopulos}}, \bibinfo {author} {\bibfnamefont {V.}~\bibnamefont
  {Karanikolas}}, \bibinfo {author} {\bibfnamefont {N.}~\bibnamefont
  {Iliopoulos}},\ and\ \bibinfo {author} {\bibfnamefont {E.}~\bibnamefont
  {Paspalakis}},\ }\bibfield  {title} {\bibinfo {title} {Non-markovian
  spontaneous emission dynamics of a quantum emitter near a
  ${\mathrm{mos}}_{2}$ nanodisk},\ }\href
  {https://doi.org/10.1103/PhysRevB.99.195412} {\bibfield  {journal} {\bibinfo
  {journal} {Phys. Rev. B}\ }\textbf {\bibinfo {volume} {99}},\ \bibinfo
  {pages} {195412} (\bibinfo {year} {2019})}\BibitemShut {NoStop}%
\bibitem [{\citenamefont {Whisler}\ \emph {et~al.}(2023)\citenamefont
  {Whisler}, \citenamefont {Holdman}, \citenamefont {Yavuz},\ and\
  \citenamefont {Brar}}]{Whisler}%
  \BibitemOpen
  \bibfield  {author} {\bibinfo {author} {\bibfnamefont {C.}~\bibnamefont
  {Whisler}}, \bibinfo {author} {\bibfnamefont {G.}~\bibnamefont {Holdman}},
  \bibinfo {author} {\bibfnamefont {D.~D.}\ \bibnamefont {Yavuz}},\ and\
  \bibinfo {author} {\bibfnamefont {V.~W.}\ \bibnamefont {Brar}},\ }\bibfield
  {title} {\bibinfo {title} {Enhancing two-photon spontaneous emission in rare
  earths using graphene and graphene nanoribbons},\ }\href
  {https://doi.org/10.1103/PhysRevB.107.195420} {\bibfield  {journal} {\bibinfo
   {journal} {Phys. Rev. B}\ }\textbf {\bibinfo {volume} {107}},\ \bibinfo
  {pages} {195420} (\bibinfo {year} {2023})}\BibitemShut {NoStop}%
\bibitem [{\citenamefont {Agarwal}(1991)}]{Agarwal91}%
  \BibitemOpen
  \bibfield  {author} {\bibinfo {author} {\bibfnamefont {G.~S.}\ \bibnamefont
  {Agarwal}},\ }\bibfield  {title} {\bibinfo {title} {Inhibition of spontaneous
  emission noise in lasers without inversion},\ }\href
  {https://doi.org/10.1103/PhysRevLett.67.980} {\bibfield  {journal} {\bibinfo
  {journal} {Phys. Rev. Lett.}\ }\textbf {\bibinfo {volume} {67}},\ \bibinfo
  {pages} {980} (\bibinfo {year} {1991})}\BibitemShut {NoStop}%
\bibitem [{\citenamefont {Kozlov}\ \emph {et~al.}(2006)\citenamefont {Kozlov},
  \citenamefont {Rostovtsev},\ and\ \citenamefont {Scully}}]{Kozlov}%
  \BibitemOpen
  \bibfield  {author} {\bibinfo {author} {\bibfnamefont {V.~V.}\ \bibnamefont
  {Kozlov}}, \bibinfo {author} {\bibfnamefont {Y.}~\bibnamefont {Rostovtsev}},\
  and\ \bibinfo {author} {\bibfnamefont {M.~O.}\ \bibnamefont {Scully}},\
  }\bibfield  {title} {\bibinfo {title} {Inducing quantum coherence via decays
  and incoherent pumping with application to population trapping, lasing
  without inversion, and quenching of spontaneous emission},\ }\href
  {https://doi.org/10.1103/PhysRevA.74.063829} {\bibfield  {journal} {\bibinfo
  {journal} {Phys. Rev. A}\ }\textbf {\bibinfo {volume} {74}},\ \bibinfo
  {pages} {063829} (\bibinfo {year} {2006})}\BibitemShut {NoStop}%
\bibitem [{\citenamefont {Xie}\ \emph {et~al.}(2012)\citenamefont {Xie},
  \citenamefont {Cai},\ and\ \citenamefont {Xiao}}]{Bin}%
  \BibitemOpen
  \bibfield  {author} {\bibinfo {author} {\bibfnamefont {B.}~\bibnamefont
  {Xie}}, \bibinfo {author} {\bibfnamefont {X.}~\bibnamefont {Cai}},\ and\
  \bibinfo {author} {\bibfnamefont {Z.-H.}\ \bibnamefont {Xiao}},\ }\bibfield
  {title} {\bibinfo {title} {Electromagnetically induced phase grating
  controlled by spontaneous emission},\ }\href@noop {} {\bibfield  {journal}
  {\bibinfo  {journal} {Optics Communications}\ }\textbf {\bibinfo {volume}
  {285}},\ \bibinfo {pages} {133} (\bibinfo {year} {2012})}\BibitemShut
  {NoStop}%
\bibitem [{\citenamefont {Bozorgzadeh}\ \emph {et~al.}(2016)\citenamefont
  {Bozorgzadeh}, \citenamefont {Sahrai},\ and\ \citenamefont
  {Khoshsima}}]{Forough}%
  \BibitemOpen
  \bibfield  {author} {\bibinfo {author} {\bibfnamefont {F.}~\bibnamefont
  {Bozorgzadeh}}, \bibinfo {author} {\bibfnamefont {M.}~\bibnamefont
  {Sahrai}},\ and\ \bibinfo {author} {\bibfnamefont {H.}~\bibnamefont
  {Khoshsima}},\ }\bibfield  {title} {\bibinfo {title} {Controlling the
  electromagnetically induced grating via spontaneously generated coherence},\
  }\href@noop {} {\bibfield  {journal} {\bibinfo  {journal} {Eur. Phys. J. D}\
  }\textbf {\bibinfo {volume} {70}},\ \bibinfo {pages} {191} (\bibinfo {year}
  {2016})}\BibitemShut {NoStop}%
\bibitem [{\citenamefont {Scully}\ and\ \citenamefont
  {Fleischhauer}(1992)}]{Scullymagnetometer}%
  \BibitemOpen
  \bibfield  {author} {\bibinfo {author} {\bibfnamefont {M.~O.}\ \bibnamefont
  {Scully}}\ and\ \bibinfo {author} {\bibfnamefont {M.}~\bibnamefont
  {Fleischhauer}},\ }\bibfield  {title} {\bibinfo {title} {High-sensitivity
  magnetometer based on index-enhanced media},\ }\href
  {https://doi.org/10.1103/PhysRevLett.69.1360} {\bibfield  {journal} {\bibinfo
   {journal} {Phys. Rev. Lett.}\ }\textbf {\bibinfo {volume} {69}},\ \bibinfo
  {pages} {1360} (\bibinfo {year} {1992})}\BibitemShut {NoStop}%
\bibitem [{\citenamefont {Wan}\ and\ \citenamefont {Zhang}(2011)}]{Ren}%
  \BibitemOpen
  \bibfield  {author} {\bibinfo {author} {\bibfnamefont {R.-G.}\ \bibnamefont
  {Wan}}\ and\ \bibinfo {author} {\bibfnamefont {T.-Y.}\ \bibnamefont
  {Zhang}},\ }\bibfield  {title} {\bibinfo {title} {Two-dimensional
  sub-half-wavelength atom localization via controlled spontaneous emission},\
  }\href@noop {} {\bibfield  {journal} {\bibinfo  {journal} {Optics Express}\
  }\textbf {\bibinfo {volume} {19}},\ \bibinfo {pages} {25823} (\bibinfo {year}
  {2011})}\BibitemShut {NoStop}%
\bibitem [{\citenamefont {Wang}\ \emph {et~al.}(2017)\citenamefont {Wang},
  \citenamefont {Chen},\ and\ \citenamefont {Yu}}]{Zhiping}%
  \BibitemOpen
  \bibfield  {author} {\bibinfo {author} {\bibfnamefont {Z.}~\bibnamefont
  {Wang}}, \bibinfo {author} {\bibfnamefont {J.}~\bibnamefont {Chen}},\ and\
  \bibinfo {author} {\bibfnamefont {B.}~\bibnamefont {Yu}},\ }\bibfield
  {title} {\bibinfo {title} {High-dimensional atom localization via
  spontaneously generated coherence in a microwave-driven atomic system},\
  }\href@noop {} {\bibfield  {journal} {\bibinfo  {journal} {Optics Express}\
  }\textbf {\bibinfo {volume} {25}},\ \bibinfo {pages} {3358} (\bibinfo {year}
  {2017})}\BibitemShut {NoStop}%
\bibitem [{\citenamefont {Scully}(1991)}]{Scully91ri}%
  \BibitemOpen
  \bibfield  {author} {\bibinfo {author} {\bibfnamefont {M.~O.}\ \bibnamefont
  {Scully}},\ }\bibfield  {title} {\bibinfo {title} {Enhancement of the index
  of refraction via quantum coherence},\ }\href
  {https://doi.org/10.1103/PhysRevLett.67.1855} {\bibfield  {journal} {\bibinfo
   {journal} {Phys. Rev. Lett.}\ }\textbf {\bibinfo {volume} {67}},\ \bibinfo
  {pages} {1855} (\bibinfo {year} {1991})}\BibitemShut {NoStop}%
\bibitem [{\citenamefont {Evangelou}\ \emph {et~al.}(2012)\citenamefont
  {Evangelou}, \citenamefont {Yannopapas},\ and\ \citenamefont
  {Paspalakis}}]{Evangelou}%
  \BibitemOpen
  \bibfield  {author} {\bibinfo {author} {\bibfnamefont {S.}~\bibnamefont
  {Evangelou}}, \bibinfo {author} {\bibfnamefont {V.}~\bibnamefont
  {Yannopapas}},\ and\ \bibinfo {author} {\bibfnamefont {E.}~\bibnamefont
  {Paspalakis}},\ }\bibfield  {title} {\bibinfo {title} {Transparency and slow
  light in a four-level quantum system near a plasmonic nanostructure},\ }\href
  {https://doi.org/10.1103/PhysRevA.86.053811} {\bibfield  {journal} {\bibinfo
  {journal} {Phys. Rev. A}\ }\textbf {\bibinfo {volume} {86}},\ \bibinfo
  {pages} {053811} (\bibinfo {year} {2012})}\BibitemShut {NoStop}%
\bibitem [{\citenamefont {Bennett}\ and\ \citenamefont
  {DiVincenzo}(2000)}]{Charles}%
  \BibitemOpen
  \bibfield  {author} {\bibinfo {author} {\bibfnamefont {C.~H.}\ \bibnamefont
  {Bennett}}\ and\ \bibinfo {author} {\bibfnamefont {D.~P.}\ \bibnamefont
  {DiVincenzo}},\ }\bibfield  {title} {\bibinfo {title} {Quantum information
  and computation},\ }\href@noop {} {\bibfield  {journal} {\bibinfo  {journal}
  {Nature}\ }\textbf {\bibinfo {volume} {404}},\ \bibinfo {pages} {247}
  (\bibinfo {year} {2000})}\BibitemShut {NoStop}%
\bibitem [{\citenamefont {Paternostro}\ \emph {et~al.}(2005)\citenamefont
  {Paternostro}, \citenamefont {Kim},\ and\ \citenamefont
  {Knight}}]{Paternostro}%
  \BibitemOpen
  \bibfield  {author} {\bibinfo {author} {\bibfnamefont {M.}~\bibnamefont
  {Paternostro}}, \bibinfo {author} {\bibfnamefont {M.~S.}\ \bibnamefont
  {Kim}},\ and\ \bibinfo {author} {\bibfnamefont {P.~L.}\ \bibnamefont
  {Knight}},\ }\bibfield  {title} {\bibinfo {title} {Vibrational coherent
  quantum computation},\ }\href {https://doi.org/10.1103/PhysRevA.71.022311}
  {\bibfield  {journal} {\bibinfo  {journal} {Phys. Rev. A}\ }\textbf {\bibinfo
  {volume} {71}},\ \bibinfo {pages} {022311} (\bibinfo {year}
  {2005})}\BibitemShut {NoStop}%
\bibitem [{\citenamefont {Zhu}\ and\ \citenamefont {Scully}(1996)}]{Zhu1996}%
  \BibitemOpen
  \bibfield  {author} {\bibinfo {author} {\bibfnamefont {S.-Y.}\ \bibnamefont
  {Zhu}}\ and\ \bibinfo {author} {\bibfnamefont {M.~O.}\ \bibnamefont
  {Scully}},\ }\bibfield  {title} {\bibinfo {title} {Spectral line elimination
  and spontaneous emission cancellation via quantum interference},\ }\href
  {https://doi.org/10.1103/PhysRevLett.76.388} {\bibfield  {journal} {\bibinfo
  {journal} {Phys. Rev. Lett.}\ }\textbf {\bibinfo {volume} {76}},\ \bibinfo
  {pages} {388} (\bibinfo {year} {1996})}\BibitemShut {NoStop}%
\bibitem [{\citenamefont {Kapale}\ \emph {et~al.}(2003)\citenamefont {Kapale},
  \citenamefont {Scully}, \citenamefont {Zhu},\ and\ \citenamefont
  {Zubairy}}]{Kapale}%
  \BibitemOpen
  \bibfield  {author} {\bibinfo {author} {\bibfnamefont {K.~T.}\ \bibnamefont
  {Kapale}}, \bibinfo {author} {\bibfnamefont {M.~O.}\ \bibnamefont {Scully}},
  \bibinfo {author} {\bibfnamefont {S.-Y.}\ \bibnamefont {Zhu}},\ and\ \bibinfo
  {author} {\bibfnamefont {M.~S.}\ \bibnamefont {Zubairy}},\ }\bibfield
  {title} {\bibinfo {title} {Quenching of spontaneous emission through
  interference of incoherent pump processes},\ }\href
  {https://doi.org/10.1103/PhysRevA.67.023804} {\bibfield  {journal} {\bibinfo
  {journal} {Phys. Rev. A}\ }\textbf {\bibinfo {volume} {67}},\ \bibinfo
  {pages} {023804} (\bibinfo {year} {2003})}\BibitemShut {NoStop}%
\bibitem [{\citenamefont {Bojer}\ \emph {et~al.}(2022)\citenamefont {Bojer},
  \citenamefont {G\"otzend\"orfer}, \citenamefont {Bachelard},\ and\
  \citenamefont {von Zanthier}}]{Bojer}%
  \BibitemOpen
  \bibfield  {author} {\bibinfo {author} {\bibfnamefont {M.}~\bibnamefont
  {Bojer}}, \bibinfo {author} {\bibfnamefont {L.}~\bibnamefont
  {G\"otzend\"orfer}}, \bibinfo {author} {\bibfnamefont {R.}~\bibnamefont
  {Bachelard}},\ and\ \bibinfo {author} {\bibfnamefont {J.}~\bibnamefont {von
  Zanthier}},\ }\bibfield  {title} {\bibinfo {title} {Engineering of
  spontaneous emission in free space via conditional measurements},\ }\href
  {https://doi.org/10.1103/PhysRevResearch.4.043022} {\bibfield  {journal}
  {\bibinfo  {journal} {Phys. Rev. Res.}\ }\textbf {\bibinfo {volume} {4}},\
  \bibinfo {pages} {043022} (\bibinfo {year} {2022})}\BibitemShut {NoStop}%
\bibitem [{\citenamefont {Evangelou}\ \emph {et~al.}(2011)\citenamefont
  {Evangelou}, \citenamefont {Yannopapas},\ and\ \citenamefont
  {Paspalakis}}]{Evangelou2011}%
  \BibitemOpen
  \bibfield  {author} {\bibinfo {author} {\bibfnamefont {S.}~\bibnamefont
  {Evangelou}}, \bibinfo {author} {\bibfnamefont {V.}~\bibnamefont
  {Yannopapas}},\ and\ \bibinfo {author} {\bibfnamefont {E.}~\bibnamefont
  {Paspalakis}},\ }\bibfield  {title} {\bibinfo {title} {Modifying free-space
  spontaneous emission near a plasmonic nanostructure},\ }\href
  {https://doi.org/10.1103/PhysRevA.83.023819} {\bibfield  {journal} {\bibinfo
  {journal} {Phys. Rev. A}\ }\textbf {\bibinfo {volume} {83}},\ \bibinfo
  {pages} {023819} (\bibinfo {year} {2011})}\BibitemShut {NoStop}%
\bibitem [{\citenamefont {Zhang}\ \emph {et~al.}(2002)\citenamefont {Zhang},
  \citenamefont {Tang}, \citenamefont {Dong},\ and\ \citenamefont
  {He}}]{Zhang2002}%
  \BibitemOpen
  \bibfield  {author} {\bibinfo {author} {\bibfnamefont {H.~Z.}\ \bibnamefont
  {Zhang}}, \bibinfo {author} {\bibfnamefont {S.~H.}\ \bibnamefont {Tang}},
  \bibinfo {author} {\bibfnamefont {P.}~\bibnamefont {Dong}},\ and\ \bibinfo
  {author} {\bibfnamefont {J.}~\bibnamefont {He}},\ }\bibfield  {title}
  {\bibinfo {title} {Quantum interference in spontaneous emission of an atom
  embedded in a double-band photonic crystal},\ }\href
  {https://doi.org/10.1103/PhysRevA.65.063802} {\bibfield  {journal} {\bibinfo
  {journal} {Phys. Rev. A}\ }\textbf {\bibinfo {volume} {65}},\ \bibinfo
  {pages} {063802} (\bibinfo {year} {2002})}\BibitemShut {NoStop}%
\bibitem [{\citenamefont {Agarwal}\ and\ \citenamefont
  {Pathak}(2004)}]{Pathak}%
  \BibitemOpen
  \bibfield  {author} {\bibinfo {author} {\bibfnamefont {G.~S.}\ \bibnamefont
  {Agarwal}}\ and\ \bibinfo {author} {\bibfnamefont {P.~K.}\ \bibnamefont
  {Pathak}},\ }\bibfield  {title} {\bibinfo {title} {dc-field-induced
  enhancement and inhibition of spontaneous emission in a cavity},\ }\href
  {https://doi.org/10.1103/PhysRevA.70.025802} {\bibfield  {journal} {\bibinfo
  {journal} {Phys. Rev. A}\ }\textbf {\bibinfo {volume} {70}},\ \bibinfo
  {pages} {025802} (\bibinfo {year} {2004})}\BibitemShut {NoStop}%
\bibitem [{\citenamefont {Allen}\ \emph {et~al.}(1992)\citenamefont {Allen},
  \citenamefont {Beijersbergen}, \citenamefont {Spreeuw},\ and\ \citenamefont
  {Woerdman}}]{Allen}%
  \BibitemOpen
  \bibfield  {author} {\bibinfo {author} {\bibfnamefont {L.}~\bibnamefont
  {Allen}}, \bibinfo {author} {\bibfnamefont {M.~W.}\ \bibnamefont
  {Beijersbergen}}, \bibinfo {author} {\bibfnamefont {R.~J.~C.}\ \bibnamefont
  {Spreeuw}},\ and\ \bibinfo {author} {\bibfnamefont {J.~P.}\ \bibnamefont
  {Woerdman}},\ }\bibfield  {title} {\bibinfo {title} {Orbital angular momentum
  of light and the transformation of laguerre-gaussian laser modes},\ }\href
  {https://doi.org/10.1103/PhysRevA.45.8185} {\bibfield  {journal} {\bibinfo
  {journal} {Phys. Rev. A}\ }\textbf {\bibinfo {volume} {45}},\ \bibinfo
  {pages} {8185} (\bibinfo {year} {1992})}\BibitemShut {NoStop}%
\bibitem [{\citenamefont {Franke-Arnold}(2022)}]{Arnold}%
  \BibitemOpen
  \bibfield  {author} {\bibinfo {author} {\bibfnamefont {S.}~\bibnamefont
  {Franke-Arnold}},\ }\bibfield  {title} {\bibinfo {title} {30 years of orbital
  angular momentum of light},\ }\href@noop {} {\bibfield  {journal} {\bibinfo
  {journal} {Nature Reviews Physics}\ }\textbf {\bibinfo {volume} {4}},\
  \bibinfo {pages} {361} (\bibinfo {year} {2022})}\BibitemShut {NoStop}%
\bibitem [{\citenamefont {Wang}\ \emph {et~al.}(2020)\citenamefont {Wang},
  \citenamefont {Castellucci},\ and\ \citenamefont {Franke-Arnold}}]{Jinwen}%
  \BibitemOpen
  \bibfield  {author} {\bibinfo {author} {\bibfnamefont {J.}~\bibnamefont
  {Wang}}, \bibinfo {author} {\bibfnamefont {F.}~\bibnamefont {Castellucci}},\
  and\ \bibinfo {author} {\bibfnamefont {S.}~\bibnamefont {Franke-Arnold}},\
  }\bibfield  {title} {\bibinfo {title} {Vectorial light–matter interaction:
  Exploring spatially structured complex light fields},\ }\href@noop {}
  {\bibfield  {journal} {\bibinfo  {journal} {AVS Quantum Sci.}\ }\textbf
  {\bibinfo {volume} {2}},\ \bibinfo {pages} {031702} (\bibinfo {year}
  {2020})}\BibitemShut {NoStop}%
\bibitem [{\citenamefont {Stevenson}\ \emph {et~al.}(2010)\citenamefont
  {Stevenson}, \citenamefont {Moore},\ and\ \citenamefont
  {Dholakia}}]{Stevenson}%
  \BibitemOpen
  \bibfield  {author} {\bibinfo {author} {\bibfnamefont {D.~J.}\ \bibnamefont
  {Stevenson}}, \bibinfo {author} {\bibfnamefont {F.~G.}\ \bibnamefont
  {Moore}},\ and\ \bibinfo {author} {\bibfnamefont {K.}~\bibnamefont
  {Dholakia}},\ }\bibfield  {title} {\bibinfo {title} {Light forces the pace:
  optical manipulation for biophotonics},\ }\href@noop {} {\bibfield  {journal}
  {\bibinfo  {journal} {J. Biomed. Opt.}\ }\textbf {\bibinfo {volume} {15}},\
  \bibinfo {pages} {041503} (\bibinfo {year} {2010})}\BibitemShut {NoStop}%
\bibitem [{\citenamefont {Macdonald}\ \emph {et~al.}(2002)\citenamefont
  {Macdonald}, \citenamefont {Paterson}, \citenamefont {Sepulveda},
  \citenamefont {Arlt}, \citenamefont {Sibbett},\ and\ \citenamefont
  {K.Dholakia}}]{Macdonald}%
  \BibitemOpen
  \bibfield  {author} {\bibinfo {author} {\bibfnamefont {M.~P.}\ \bibnamefont
  {Macdonald}}, \bibinfo {author} {\bibfnamefont {L.}~\bibnamefont {Paterson}},
  \bibinfo {author} {\bibfnamefont {K.~V.}\ \bibnamefont {Sepulveda}}, \bibinfo
  {author} {\bibfnamefont {J.}~\bibnamefont {Arlt}}, \bibinfo {author}
  {\bibfnamefont {W.}~\bibnamefont {Sibbett}},\ and\ \bibinfo {author}
  {\bibnamefont {K.Dholakia}},\ }\bibfield  {title} {\bibinfo {title} {Creation
  and manipulation of three-dimensional optically trapped structures},\
  }\href@noop {} {\bibfield  {journal} {\bibinfo  {journal} {Science}\ }\textbf
  {\bibinfo {volume} {296}},\ \bibinfo {pages} {1101} (\bibinfo {year}
  {2002})}\BibitemShut {NoStop}%
\bibitem [{\citenamefont {Woerdemann}\ \emph {et~al.}(2013)\citenamefont
  {Woerdemann}, \citenamefont {Alpmann}, \citenamefont {Esseling},\ and\
  \citenamefont {Denz}}]{Woerdemann}%
  \BibitemOpen
  \bibfield  {author} {\bibinfo {author} {\bibfnamefont {M.}~\bibnamefont
  {Woerdemann}}, \bibinfo {author} {\bibfnamefont {C.}~\bibnamefont {Alpmann}},
  \bibinfo {author} {\bibfnamefont {M.}~\bibnamefont {Esseling}},\ and\
  \bibinfo {author} {\bibfnamefont {C.}~\bibnamefont {Denz}},\ }\bibfield
  {title} {\bibinfo {title} {Advanced optical trapping by complex beam
  shaping},\ }\href@noop {} {\bibfield  {journal} {\bibinfo  {journal} {Laser
  Photon.Rev.}\ }\textbf {\bibinfo {volume} {7}},\ \bibinfo {pages} {839}
  (\bibinfo {year} {2013})}\BibitemShut {NoStop}%
\bibitem [{\citenamefont {Dutton}\ and\ \citenamefont
  {Ruostekoski}(2004)}]{Dutton}%
  \BibitemOpen
  \bibfield  {author} {\bibinfo {author} {\bibfnamefont {Z.}~\bibnamefont
  {Dutton}}\ and\ \bibinfo {author} {\bibfnamefont {J.}~\bibnamefont
  {Ruostekoski}},\ }\bibfield  {title} {\bibinfo {title} {Transfer and storage
  of vortex states in light and matter waves},\ }\href
  {https://doi.org/10.1103/PhysRevLett.93.193602} {\bibfield  {journal}
  {\bibinfo  {journal} {Phys. Rev. Lett.}\ }\textbf {\bibinfo {volume} {93}},\
  \bibinfo {pages} {193602} (\bibinfo {year} {2004})}\BibitemShut {NoStop}%
\bibitem [{\citenamefont {Chen}\ \emph {et~al.}(2008)\citenamefont {Chen},
  \citenamefont {Shi}, \citenamefont {Zhang},\ and\ \citenamefont
  {Guo}}]{Chen2008}%
  \BibitemOpen
  \bibfield  {author} {\bibinfo {author} {\bibfnamefont {Q.-F.}\ \bibnamefont
  {Chen}}, \bibinfo {author} {\bibfnamefont {B.-S.}\ \bibnamefont {Shi}},
  \bibinfo {author} {\bibfnamefont {Y.-S.}\ \bibnamefont {Zhang}},\ and\
  \bibinfo {author} {\bibfnamefont {G.-C.}\ \bibnamefont {Guo}},\ }\bibfield
  {title} {\bibinfo {title} {Entanglement of the orbital angular momentum
  states of the photon pairs generated in a hot atomic ensemble},\ }\href
  {https://doi.org/10.1103/PhysRevA.78.053810} {\bibfield  {journal} {\bibinfo
  {journal} {Phys. Rev. A}\ }\textbf {\bibinfo {volume} {78}},\ \bibinfo
  {pages} {053810} (\bibinfo {year} {2008})}\BibitemShut {NoStop}%
\bibitem [{\citenamefont {Ruseckas}\ \emph {et~al.}(2007)\citenamefont
  {Ruseckas}, \citenamefont {Juzeli\ifmmode~\bar{u}\else \={u}\fi{}nas},
  \citenamefont {\"Ohberg},\ and\ \citenamefont {Barnett}}]{Ruseckas1}%
  \BibitemOpen
  \bibfield  {author} {\bibinfo {author} {\bibfnamefont {J.}~\bibnamefont
  {Ruseckas}}, \bibinfo {author} {\bibfnamefont {G.}~\bibnamefont
  {Juzeli\ifmmode~\bar{u}\else \={u}\fi{}nas}}, \bibinfo {author}
  {\bibfnamefont {P.}~\bibnamefont {\"Ohberg}},\ and\ \bibinfo {author}
  {\bibfnamefont {S.~M.}\ \bibnamefont {Barnett}},\ }\bibfield  {title}
  {\bibinfo {title} {Polarization rotation of slow light with orbital angular
  momentum in ultracold atomic gases},\ }\href
  {https://doi.org/10.1103/PhysRevA.76.053822} {\bibfield  {journal} {\bibinfo
  {journal} {Phys. Rev. A}\ }\textbf {\bibinfo {volume} {76}},\ \bibinfo
  {pages} {053822} (\bibinfo {year} {2007})}\BibitemShut {NoStop}%
\bibitem [{\citenamefont {Lembessis}\ and\ \citenamefont
  {Babiker}(2010)}]{Lembessis}%
  \BibitemOpen
  \bibfield  {author} {\bibinfo {author} {\bibfnamefont {V.~E.}\ \bibnamefont
  {Lembessis}}\ and\ \bibinfo {author} {\bibfnamefont {M.}~\bibnamefont
  {Babiker}},\ }\bibfield  {title} {\bibinfo {title} {Light-induced torque for
  the generation of persistent current flow in atomic gas bose-einstein
  condensates},\ }\href {https://doi.org/10.1103/PhysRevA.82.051402} {\bibfield
   {journal} {\bibinfo  {journal} {Phys. Rev. A}\ }\textbf {\bibinfo {volume}
  {82}},\ \bibinfo {pages} {051402} (\bibinfo {year} {2010})}\BibitemShut
  {NoStop}%
\bibitem [{\citenamefont {Ding}\ \emph {et~al.}(2012)\citenamefont {Ding},
  \citenamefont {Zhou}, \citenamefont {Shi}, \citenamefont {Zou},\ and\
  \citenamefont {Guo}}]{Ding}%
  \BibitemOpen
  \bibfield  {author} {\bibinfo {author} {\bibfnamefont {D.-S.}\ \bibnamefont
  {Ding}}, \bibinfo {author} {\bibfnamefont {Z.-Y.}\ \bibnamefont {Zhou}},
  \bibinfo {author} {\bibfnamefont {B.-S.}\ \bibnamefont {Shi}}, \bibinfo
  {author} {\bibfnamefont {X.-B.}\ \bibnamefont {Zou}},\ and\ \bibinfo {author}
  {\bibfnamefont {G.-C.}\ \bibnamefont {Guo}},\ }\bibfield  {title} {\bibinfo
  {title} {Linear up-conversion of orbital angular momentum},\ }\href@noop {}
  {\bibfield  {journal} {\bibinfo  {journal} {Optics Letters}\ }\textbf
  {\bibinfo {volume} {37}},\ \bibinfo {pages} {3270} (\bibinfo {year}
  {2012})}\BibitemShut {NoStop}%
\bibitem [{\citenamefont {Ruseckas}\ \emph {et~al.}(2013)\citenamefont
  {Ruseckas}, \citenamefont {Kudria\ifmmode~\check{s}\else \v{s}\fi{}ov},
  \citenamefont {Yu},\ and\ \citenamefont {Juzeli\ifmmode~\bar{u}\else
  \={u}\fi{}nas}}]{Ruseckas2}%
  \BibitemOpen
  \bibfield  {author} {\bibinfo {author} {\bibfnamefont {J.}~\bibnamefont
  {Ruseckas}}, \bibinfo {author} {\bibfnamefont {V.~c.~v.}\ \bibnamefont
  {Kudria\ifmmode~\check{s}\else \v{s}\fi{}ov}}, \bibinfo {author}
  {\bibfnamefont {I.~A.}\ \bibnamefont {Yu}},\ and\ \bibinfo {author}
  {\bibfnamefont {G.}~\bibnamefont {Juzeli\ifmmode~\bar{u}\else
  \={u}\fi{}nas}},\ }\bibfield  {title} {\bibinfo {title} {Transfer of orbital
  angular momentum of light using two-component slow light},\ }\href
  {https://doi.org/10.1103/PhysRevA.87.053840} {\bibfield  {journal} {\bibinfo
  {journal} {Phys. Rev. A}\ }\textbf {\bibinfo {volume} {87}},\ \bibinfo
  {pages} {053840} (\bibinfo {year} {2013})}\BibitemShut {NoStop}%
\bibitem [{\citenamefont {Radwell}\ \emph {et~al.}(2015)\citenamefont
  {Radwell}, \citenamefont {Clark}, \citenamefont {Piccirillo}, \citenamefont
  {Barnett},\ and\ \citenamefont {Franke-Arnold}}]{Radwell}%
  \BibitemOpen
  \bibfield  {author} {\bibinfo {author} {\bibfnamefont {N.}~\bibnamefont
  {Radwell}}, \bibinfo {author} {\bibfnamefont {T.~W.}\ \bibnamefont {Clark}},
  \bibinfo {author} {\bibfnamefont {B.}~\bibnamefont {Piccirillo}}, \bibinfo
  {author} {\bibfnamefont {S.~M.}\ \bibnamefont {Barnett}},\ and\ \bibinfo
  {author} {\bibfnamefont {S.}~\bibnamefont {Franke-Arnold}},\ }\bibfield
  {title} {\bibinfo {title} {Spatially dependent electromagnetically induced
  transparency},\ }\href {https://doi.org/10.1103/PhysRevLett.114.123603}
  {\bibfield  {journal} {\bibinfo  {journal} {Phys. Rev. Lett.}\ }\textbf
  {\bibinfo {volume} {114}},\ \bibinfo {pages} {123603} (\bibinfo {year}
  {2015})}\BibitemShut {NoStop}%
\bibitem [{\citenamefont {Sharma}\ and\ \citenamefont {Dey}(2017)}]{Sharma}%
  \BibitemOpen
  \bibfield  {author} {\bibinfo {author} {\bibfnamefont {S.}~\bibnamefont
  {Sharma}}\ and\ \bibinfo {author} {\bibfnamefont {T.~N.}\ \bibnamefont
  {Dey}},\ }\bibfield  {title} {\bibinfo {title} {Phase-induced
  transparency-mediated structured-beam generation in a closed-loop tripod
  configuration},\ }\href {https://doi.org/10.1103/PhysRevA.96.033811}
  {\bibfield  {journal} {\bibinfo  {journal} {Phys. Rev. A}\ }\textbf {\bibinfo
  {volume} {96}},\ \bibinfo {pages} {033811} (\bibinfo {year}
  {2017})}\BibitemShut {NoStop}%
\bibitem [{\citenamefont {R.Hamedi}\ \emph {et~al.}(2018)\citenamefont
  {R.Hamedi}, \citenamefont {Kudriasov}, \citenamefont {Ruseckas},\ and\
  \citenamefont {Juzeliunas}}]{HamidOE1}%
  \BibitemOpen
  \bibfield  {author} {\bibinfo {author} {\bibfnamefont {H.}~\bibnamefont
  {R.Hamedi}}, \bibinfo {author} {\bibfnamefont {V.}~\bibnamefont {Kudriasov}},
  \bibinfo {author} {\bibfnamefont {J.}~\bibnamefont {Ruseckas}},\ and\
  \bibinfo {author} {\bibfnamefont {G.}~\bibnamefont {Juzeliunas}},\ }\bibfield
   {title} {\bibinfo {title} {Azimuthal modulation of electromagnetically
  induced transparency using structured light},\ }\href@noop {} {\bibfield
  {journal} {\bibinfo  {journal} {Optics Express}\ }\textbf {\bibinfo {volume}
  {26}},\ \bibinfo {pages} {28249} (\bibinfo {year} {2018})}\BibitemShut
  {NoStop}%
\bibitem [{\citenamefont {Hamedi}\ \emph {et~al.}(2020)\citenamefont {Hamedi},
  \citenamefont {Ruseckas}, \citenamefont {Paspalakis},\ and\ \citenamefont
  {Juzeli\ifmmode~\bar{u}\else \={u}\fi{}nas}}]{HamediOFF}%
  \BibitemOpen
  \bibfield  {author} {\bibinfo {author} {\bibfnamefont {H.~R.}\ \bibnamefont
  {Hamedi}}, \bibinfo {author} {\bibfnamefont {J.}~\bibnamefont {Ruseckas}},
  \bibinfo {author} {\bibfnamefont {E.}~\bibnamefont {Paspalakis}},\ and\
  \bibinfo {author} {\bibfnamefont {G.}~\bibnamefont
  {Juzeli\ifmmode~\bar{u}\else \={u}\fi{}nas}},\ }\bibfield  {title} {\bibinfo
  {title} {Off-axis optical vortices using double-raman singlet and doublet
  light-matter schemes},\ }\href {https://doi.org/10.1103/PhysRevA.101.063828}
  {\bibfield  {journal} {\bibinfo  {journal} {Phys. Rev. A}\ }\textbf {\bibinfo
  {volume} {101}},\ \bibinfo {pages} {063828} (\bibinfo {year}
  {2020})}\BibitemShut {NoStop}%
\bibitem [{\citenamefont {Qiu}\ \emph {et~al.}(2020)\citenamefont {Qiu},
  \citenamefont {Wang}, \citenamefont {Ding}, \citenamefont {Li},\ and\
  \citenamefont {Yu}}]{Jing}%
  \BibitemOpen
  \bibfield  {author} {\bibinfo {author} {\bibfnamefont {J.}~\bibnamefont
  {Qiu}}, \bibinfo {author} {\bibfnamefont {Z.}~\bibnamefont {Wang}}, \bibinfo
  {author} {\bibfnamefont {D.}~\bibnamefont {Ding}}, \bibinfo {author}
  {\bibfnamefont {W.}~\bibnamefont {Li}},\ and\ \bibinfo {author}
  {\bibfnamefont {B.}~\bibnamefont {Yu}},\ }\bibfield  {title} {\bibinfo
  {title} {Highly efficient vortex four-wave mixing in asymmetric semiconductor
  quantum wells},\ }\href@noop {} {\bibfield  {journal} {\bibinfo  {journal}
  {Optics Express}\ }\textbf {\bibinfo {volume} {28}},\ \bibinfo {pages} {2975}
  (\bibinfo {year} {2020})}\BibitemShut {NoStop}%
\bibitem [{\citenamefont {Asadpour}\ \emph {et~al.}(2021)\citenamefont
  {Asadpour}, \citenamefont {Kirova}, \citenamefont {Qian}, \citenamefont
  {Hamedi}, \citenamefont {Juzeliunas},\ and\ \citenamefont
  {Paspalakis}}]{Seyyed}%
  \BibitemOpen
  \bibfield  {author} {\bibinfo {author} {\bibfnamefont {S.~H.}\ \bibnamefont
  {Asadpour}}, \bibinfo {author} {\bibfnamefont {T.}~\bibnamefont {Kirova}},
  \bibinfo {author} {\bibfnamefont {J.}~\bibnamefont {Qian}}, \bibinfo {author}
  {\bibfnamefont {H.~R.}\ \bibnamefont {Hamedi}}, \bibinfo {author}
  {\bibfnamefont {G.}~\bibnamefont {Juzeliunas}},\ and\ \bibinfo {author}
  {\bibfnamefont {E.}~\bibnamefont {Paspalakis}},\ }\bibfield  {title}
  {\bibinfo {title} {Azimuthal modulation of electromagnetically induced
  grating using structured light},\ }\href@noop {} {\bibfield  {journal}
  {\bibinfo  {journal} {Scientific Reports}\ }\textbf {\bibinfo {volume}
  {11}},\ \bibinfo {pages} {20721} (\bibinfo {year} {2021})}\BibitemShut
  {NoStop}%
\bibitem [{\citenamefont {Song}\ \emph {et~al.}(2022)\citenamefont {Song},
  \citenamefont {Wang}, \citenamefont {Li}, \citenamefont {Yu},\ and\
  \citenamefont {Huang}}]{Song}%
  \BibitemOpen
  \bibfield  {author} {\bibinfo {author} {\bibfnamefont {F.}~\bibnamefont
  {Song}}, \bibinfo {author} {\bibfnamefont {Z.}~\bibnamefont {Wang}}, \bibinfo
  {author} {\bibfnamefont {E.}~\bibnamefont {Li}}, \bibinfo {author}
  {\bibfnamefont {B.}~\bibnamefont {Yu}},\ and\ \bibinfo {author}
  {\bibfnamefont {Z.}~\bibnamefont {Huang}},\ }\bibfield  {title} {\bibinfo
  {title} {Nonreciprocity with structured light using optical pumping in hot
  atoms},\ }\href {https://doi.org/10.1103/PhysRevApplied.18.024027} {\bibfield
   {journal} {\bibinfo  {journal} {Phys. Rev. Appl.}\ }\textbf {\bibinfo
  {volume} {18}},\ \bibinfo {pages} {024027} (\bibinfo {year}
  {2022})}\BibitemShut {NoStop}%
\bibitem [{\citenamefont {Daloi}\ \emph {et~al.}(2022)\citenamefont {Daloi},
  \citenamefont {Kumar},\ and\ \citenamefont {Dey}}]{Daloi}%
  \BibitemOpen
  \bibfield  {author} {\bibinfo {author} {\bibfnamefont {N.}~\bibnamefont
  {Daloi}}, \bibinfo {author} {\bibfnamefont {P.}~\bibnamefont {Kumar}},\ and\
  \bibinfo {author} {\bibfnamefont {T.~N.}\ \bibnamefont {Dey}},\ }\bibfield
  {title} {\bibinfo {title} {Guiding and polarization shaping of vector beams
  in anisotropic media},\ }\href {https://doi.org/10.1103/PhysRevA.105.063714}
  {\bibfield  {journal} {\bibinfo  {journal} {Phys. Rev. A}\ }\textbf {\bibinfo
  {volume} {105}},\ \bibinfo {pages} {063714} (\bibinfo {year}
  {2022})}\BibitemShut {NoStop}%
\bibitem [{\citenamefont {Hamid R.~Hamedi}\ \emph {et~al.}(2022)\citenamefont
  {Hamid R.~Hamedi}, \citenamefont {Ahufinger}, \citenamefont {Halfmann},
  \citenamefont {Mompart},\ and\ \citenamefont {Juzeliunas}}]{HamidOEloc}%
  \BibitemOpen
  \bibfield  {author} {\bibinfo {author} {\bibfnamefont {G.~Z.}\ \bibnamefont
  {Hamid R.~Hamedi}}, \bibinfo {author} {\bibfnamefont {V.}~\bibnamefont
  {Ahufinger}}, \bibinfo {author} {\bibfnamefont {T.}~\bibnamefont {Halfmann}},
  \bibinfo {author} {\bibfnamefont {J.}~\bibnamefont {Mompart}},\ and\ \bibinfo
  {author} {\bibfnamefont {G.}~\bibnamefont {Juzeliunas}},\ }\bibfield  {title}
  {\bibinfo {title} {Spatially strongly confined atomic excitation via a two
  dimensional stimulated raman adiabatic passage},\ }\href@noop {} {\bibfield
  {journal} {\bibinfo  {journal} {Optics Express}\ }\textbf {\bibinfo {volume}
  {30}},\ \bibinfo {pages} {13915} (\bibinfo {year} {2022})}\BibitemShut
  {NoStop}%
\bibitem [{\citenamefont {Hamedi}\ \emph
  {et~al.}(2023{\natexlab{a}})\citenamefont {Hamedi}, \citenamefont {Yu},\ and\
  \citenamefont {Paspalakis}}]{Hamedimatch}%
  \BibitemOpen
  \bibfield  {author} {\bibinfo {author} {\bibfnamefont {H.~R.}\ \bibnamefont
  {Hamedi}}, \bibinfo {author} {\bibfnamefont {I.~A.}\ \bibnamefont {Yu}},\
  and\ \bibinfo {author} {\bibfnamefont {E.}~\bibnamefont {Paspalakis}},\
  }\bibfield  {title} {\bibinfo {title} {Matched optical vortices of slow light
  using a tripod coherently prepared scheme},\ }\href
  {https://doi.org/10.1103/PhysRevA.108.053719} {\bibfield  {journal} {\bibinfo
   {journal} {Phys. Rev. A}\ }\textbf {\bibinfo {volume} {108}},\ \bibinfo
  {pages} {053719} (\bibinfo {year} {2023}{\natexlab{a}})}\BibitemShut
  {NoStop}%
\bibitem [{\citenamefont {Chen}\ \emph {et~al.}(2023)\citenamefont {Chen},
  \citenamefont {Zhou}, \citenamefont {Li}, \citenamefont {Liu}, \citenamefont
  {Ciappina},\ and\ \citenamefont {Lu}}]{Chen2023}%
  \BibitemOpen
  \bibfield  {author} {\bibinfo {author} {\bibfnamefont {Y.}~\bibnamefont
  {Chen}}, \bibinfo {author} {\bibfnamefont {Y.}~\bibnamefont {Zhou}}, \bibinfo
  {author} {\bibfnamefont {M.}~\bibnamefont {Li}}, \bibinfo {author}
  {\bibfnamefont {K.}~\bibnamefont {Liu}}, \bibinfo {author} {\bibfnamefont
  {M.~F.}\ \bibnamefont {Ciappina}},\ and\ \bibinfo {author} {\bibfnamefont
  {P.}~\bibnamefont {Lu}},\ }\bibfield  {title} {\bibinfo {title} {Atomic
  photoionization by spatiotemporal optical vortex pulses},\ }\href
  {https://doi.org/10.1103/PhysRevA.107.033112} {\bibfield  {journal} {\bibinfo
   {journal} {Phys. Rev. A}\ }\textbf {\bibinfo {volume} {107}},\ \bibinfo
  {pages} {033112} (\bibinfo {year} {2023})}\BibitemShut {NoStop}%
\bibitem [{\citenamefont {Meng}\ \emph {et~al.}(2023)\citenamefont {Meng},
  \citenamefont {Shui},\ and\ \citenamefont {Yang}}]{Meng}%
  \BibitemOpen
  \bibfield  {author} {\bibinfo {author} {\bibfnamefont {C.}~\bibnamefont
  {Meng}}, \bibinfo {author} {\bibfnamefont {T.}~\bibnamefont {Shui}},\ and\
  \bibinfo {author} {\bibfnamefont {W.-X.}\ \bibnamefont {Yang}},\ }\bibfield
  {title} {\bibinfo {title} {Coherent transfer of optical vortices via backward
  four-wave mixing in a double-$\mathrm{\ensuremath{\Lambda}}$ atomic system},\
  }\href {https://doi.org/10.1103/PhysRevA.107.053712} {\bibfield  {journal}
  {\bibinfo  {journal} {Phys. Rev. A}\ }\textbf {\bibinfo {volume} {107}},\
  \bibinfo {pages} {053712} (\bibinfo {year} {2023})}\BibitemShut {NoStop}%
\bibitem [{\citenamefont {Fleischhauer}\ \emph {et~al.}(2005)\citenamefont
  {Fleischhauer}, \citenamefont {Imamoglu},\ and\ \citenamefont
  {Marangos}}]{Fleischhauer}%
  \BibitemOpen
  \bibfield  {author} {\bibinfo {author} {\bibfnamefont {M.}~\bibnamefont
  {Fleischhauer}}, \bibinfo {author} {\bibfnamefont {A.}~\bibnamefont
  {Imamoglu}},\ and\ \bibinfo {author} {\bibfnamefont {J.~P.}\ \bibnamefont
  {Marangos}},\ }\bibfield  {title} {\bibinfo {title} {Electromagnetically
  induced transparency: Optics in coherent media},\ }\href
  {https://doi.org/10.1103/RevModPhys.77.633} {\bibfield  {journal} {\bibinfo
  {journal} {Rev. Mod. Phys.}\ }\textbf {\bibinfo {volume} {77}},\ \bibinfo
  {pages} {633} (\bibinfo {year} {2005})}\BibitemShut {NoStop}%
\bibitem [{\citenamefont {Hamedi}\ \emph
  {et~al.}(2023{\natexlab{b}})\citenamefont {Hamedi}, \citenamefont
  {Yannopapas}, \citenamefont {Paspalakis},\ and\ \citenamefont
  {Ruseckas}}]{Hamid2023}%
  \BibitemOpen
  \bibfield  {author} {\bibinfo {author} {\bibfnamefont {H.~R.}\ \bibnamefont
  {Hamedi}}, \bibinfo {author} {\bibfnamefont {V.}~\bibnamefont {Yannopapas}},
  \bibinfo {author} {\bibfnamefont {E.}~\bibnamefont {Paspalakis}},\ and\
  \bibinfo {author} {\bibfnamefont {J.}~\bibnamefont {Ruseckas}},\ }\bibfield
  {title} {\bibinfo {title} {Spatially patterned light amplification without
  inversion},\ }\href@noop {} {\bibfield  {journal} {\bibinfo  {journal}
  {Results in Physics}\ }\textbf {\bibinfo {volume} {54}},\ \bibinfo {pages}
  {107135} (\bibinfo {year} {2023}{\natexlab{b}})}\BibitemShut {NoStop}%
\bibitem [{\citenamefont {Abbas}\ \emph {et~al.}(2024)\citenamefont {Abbas},
  \citenamefont {Saleem}, \citenamefont {Rahmatullah}, \citenamefont {Zhang},\
  and\ \citenamefont {Zhang}}]{Abbas2024}%
  \BibitemOpen
  \bibfield  {author} {\bibinfo {author} {\bibfnamefont {M.}~\bibnamefont
  {Abbas}}, \bibinfo {author} {\bibfnamefont {U.}~\bibnamefont {Saleem}},
  \bibinfo {author} {\bibnamefont {Rahmatullah}}, \bibinfo {author}
  {\bibfnamefont {Y.-C.}\ \bibnamefont {Zhang}},\ and\ \bibinfo {author}
  {\bibfnamefont {P.}~\bibnamefont {Zhang}},\ }\bibfield  {title} {\bibinfo
  {title} {Spontaneously generated structured light in a coherently driven
  five-level m-type atomic system},\ }\href
  {https://doi.org/10.1103/PhysRevA.109.023716} {\bibfield  {journal} {\bibinfo
   {journal} {Phys. Rev. A}\ }\textbf {\bibinfo {volume} {109}},\ \bibinfo
  {pages} {023716} (\bibinfo {year} {2024})}\BibitemShut {NoStop}%
\bibitem [{\citenamefont {Agarwal}(1974)}]{Agarwalbook}%
  \BibitemOpen
  \bibfield  {author} {\bibinfo {author} {\bibfnamefont {G.~S.}\ \bibnamefont
  {Agarwal}},\ }\href@noop {} {\emph {\bibinfo {title} {Quantum Statistical
  Theories of Spontaneous Emission and their Relation to Other Approaches}}},\
  edited by\ \bibinfo {editor} {\bibfnamefont {G.~H.}\ \bibnamefont {et~al.}},\
  Vol.~\bibinfo {volume} {70}\ (\bibinfo  {publisher} {Springer Tracks in
  Modern Physics},\ \bibinfo {year} {1974})\BibitemShut {NoStop}%
\bibitem [{\citenamefont {Barnett}\ and\ \citenamefont
  {Radmore}(1997)}]{Stephen}%
  \BibitemOpen
  \bibfield  {author} {\bibinfo {author} {\bibfnamefont {S.~M.}\ \bibnamefont
  {Barnett}}\ and\ \bibinfo {author} {\bibfnamefont {P.~M.}\ \bibnamefont
  {Radmore}},\ }\href@noop {} {\emph {\bibinfo {title} {Methods in Theoretical
  Quantum Optics}}},\ edited by\ \bibinfo {editor} {\bibfnamefont {O.~O.~U.}\
  \bibnamefont {Press}}\ (\bibinfo  {publisher} {Oxford: Oxford University
  Press},\ \bibinfo {year} {1997})\BibitemShut {NoStop}%
\bibitem [{\citenamefont {Uhlenberg}\ and\ \citenamefont
  {Uhlenberg}(2000)}]{Uhlenberg2000MagnetoopticalTO}%
  \BibitemOpen
  \bibfield  {author} {\bibinfo {author} {\bibfnamefont {G.}~\bibnamefont
  {Uhlenberg}}\ and\ \bibinfo {author} {\bibfnamefont {J.}~\bibnamefont
  {Dirscherl}} and\ \bibinfo {author} {\bibfnamefont {H.}~\bibnamefont
  {Walther}},\ }\bibfield  {title} {\bibinfo {title} {Magneto-optical trapping of silver atoms},\ }\href
  {https://api.semanticscholar.org/CorpusID:119694853} {\bibfield
  {journal} {\bibinfo  {journal} {Phys. Rev. A}\ }\textbf {\bibinfo
  {volume} {62}},\ \bibinfo {pages} {063404} (\bibinfo {year}
  {2000})}\BibitemShut {NoStop}%
\bibitem [{\citenamefont {Friederich}\ and\ \citenamefont
  {Friederich}(2010)}]{Friederich2010PhaselockingOT}%
  \BibitemOpen
  \bibfield  {author} {\bibinfo {author} {\bibfnamefont {F.}~\bibnamefont
  {Friederich}}\ and\ \bibinfo {author} {\bibfnamefont {G.}~\bibnamefont
  {Schuricht}} and\ \bibinfo {author} {\bibfnamefont {A.}~\bibnamefont
  {Deninger}} and\ \bibinfo {author} {\bibfnamefont {F.}~\bibnamefont
  {Lison}} and\ \bibinfo {author} {\bibfnamefont {G.}~\bibnamefont
  {Spickermann}} and\ \bibinfo {author} {\bibfnamefont {P.}~\bibnamefont
  {Haring}},\ }\bibfield  {title} {\bibinfo {title} {Phase-locking of the beat signal of two distributed-feedback diode lasers to oscillators working in the MHz to THz range},\ }\href
  {https://api.semanticscholar.org/CorpusID:19406141} {\bibfield
  {journal} {\bibinfo  {journal} {Optics Express}\ }\textbf {\bibinfo
  {volume} {18}},\ \bibinfo {pages} {8621} (\bibinfo {year}
  {2010})}\BibitemShut {NoStop}%
 \bibitem [{\citenamefont {Norris}\ \emph {et~al.}(2010)\citenamefont {Norris},
  \citenamefont {Orozco}, \citenamefont {Barberis-Blostein},\ and\
  \citenamefont {Carmichael}}]{Norris}%
  \BibitemOpen
  \bibfield  {author} {\bibinfo {author} {\bibfnamefont {D.~G.}\ \bibnamefont
  {Norris}}, \bibinfo {author} {\bibfnamefont {L.~A.}\ \bibnamefont {Orozco}},
  \bibinfo {author} {\bibfnamefont {P.}~\bibnamefont {Barberis-Blostein}},\
  and\ \bibinfo {author} {\bibfnamefont {H.~J.}\ \bibnamefont {Carmichael}},\
  }\bibfield  {title} {\bibinfo {title} {Observation of ground-state quantum
  beats in atomic spontaneous emission},\ }\href
  {https://doi.org/10.1103/PhysRevLett.105.123602} {\bibfield  {journal}
  {\bibinfo  {journal} {Phys. Rev. Lett.}\ }\textbf {\bibinfo {volume} {105}},\
  \bibinfo {pages} {123602} (\bibinfo {year} {2010})}\BibitemShut {NoStop}%
\bibitem [{\citenamefont {Heeg}\ \emph {et~al.}(2013)\citenamefont {Heeg},
  \citenamefont {Wille}, \citenamefont {Schlage}, \citenamefont {Guryeva},
  \citenamefont {Schumacher}, \citenamefont {Uschmann}, \citenamefont
  {Schulze}, \citenamefont {Marx}, \citenamefont {K\"ampfer}, \citenamefont
  {Paulus}, \citenamefont {R\"ohlsberger},\ and\ \citenamefont {Evers}}]{Heeg}%
  \BibitemOpen
  \bibfield  {author} {\bibinfo {author} {\bibfnamefont {K.~P.}\ \bibnamefont
  {Heeg}}, \bibinfo {author} {\bibfnamefont {H.-C.}\ \bibnamefont {Wille}},
  \bibinfo {author} {\bibfnamefont {K.}~\bibnamefont {Schlage}}, \bibinfo
  {author} {\bibfnamefont {T.}~\bibnamefont {Guryeva}}, \bibinfo {author}
  {\bibfnamefont {D.}~\bibnamefont {Schumacher}}, \bibinfo {author}
  {\bibfnamefont {I.}~\bibnamefont {Uschmann}}, \bibinfo {author}
  {\bibfnamefont {K.~S.}\ \bibnamefont {Schulze}}, \bibinfo {author}
  {\bibfnamefont {B.}~\bibnamefont {Marx}}, \bibinfo {author} {\bibfnamefont
  {T.}~\bibnamefont {K\"ampfer}}, \bibinfo {author} {\bibfnamefont {G.~G.}\
  \bibnamefont {Paulus}}, \bibinfo {author} {\bibfnamefont {R.}~\bibnamefont
  {R\"ohlsberger}},\ and\ \bibinfo {author} {\bibfnamefont {J.}~\bibnamefont
  {Evers}},\ }\bibfield  {title} {\bibinfo {title} {Vacuum-assisted generation
  and control of atomic coherences at x-ray energies},\ }\href
  {https://doi.org/10.1103/PhysRevLett.111.073601} {\bibfield  {journal}
  {\bibinfo  {journal} {Phys. Rev. Lett.}\ }\textbf {\bibinfo {volume} {111}},\
  \bibinfo {pages} {073601} (\bibinfo {year} {2013})}\BibitemShut {NoStop}%
\bibitem [{\citenamefont {Xia}\ \emph {et~al.}(1996)\citenamefont {Xia},
  \citenamefont {Ye},\ and\ \citenamefont {Zhu}}]{Xia}%
  \BibitemOpen
  \bibfield  {author} {\bibinfo {author} {\bibfnamefont {H.-R.}~\bibnamefont
  {Xia}}\ and\ \bibinfo {author} {\bibfnamefont {C.-Y.}~\bibnamefont
  {Ye}} and\ \bibinfo {author} {\bibfnamefont {S.-Y.}~\bibnamefont
  {Zhu}},\ }\bibfield  {title} {\bibinfo {title} {Experimental Observation of Spontaneous Emission Cancellation},\ }\href {https://link.aps.org/doi/10.1103/PhysRevLett.77.1032} {\bibfield
  {journal} {\bibinfo  {journal} {Phys. Rev. Lett.}\ }\textbf {\bibinfo
  {volume} {77}},\ \bibinfo {pages} {1032} (\bibinfo {year}
  {1996})}\BibitemShut {NoStop}%
\bibitem [{\citenamefont {Dutt}\ \emph {et~al.}(2005)\citenamefont {Dutt},
  \citenamefont {Cheng}, \citenamefont {Li}, \citenamefont {Xu}, \citenamefont
  {Li}, \citenamefont {Berman}, \citenamefont {Steel}, \citenamefont {Bracker},
  \citenamefont {Gammon}, \citenamefont {Economou}, \citenamefont {Liu},\ and\
  \citenamefont {Sham}}]{Dutt2005}%
  \BibitemOpen
  \bibfield  {author} {\bibinfo {author} {\bibfnamefont {M.~V.~G.}\
  \bibnamefont {Dutt}}, \bibinfo {author} {\bibfnamefont {J.}~\bibnamefont
  {Cheng}}, \bibinfo {author} {\bibfnamefont {B.}~\bibnamefont {Li}}, \bibinfo
  {author} {\bibfnamefont {X.}~\bibnamefont {Xu}}, \bibinfo {author}
  {\bibfnamefont {X.}~\bibnamefont {Li}}, \bibinfo {author} {\bibfnamefont
  {P.~R.}\ \bibnamefont {Berman}}, \bibinfo {author} {\bibfnamefont {D.~G.}\
  \bibnamefont {Steel}}, \bibinfo {author} {\bibfnamefont {A.~S.}\ \bibnamefont
  {Bracker}}, \bibinfo {author} {\bibfnamefont {D.}~\bibnamefont {Gammon}},
  \bibinfo {author} {\bibfnamefont {S.~E.}\ \bibnamefont {Economou}}, \bibinfo
  {author} {\bibfnamefont {R.-B.}\ \bibnamefont {Liu}},\ and\ \bibinfo {author}
  {\bibfnamefont {L.~J.}\ \bibnamefont {Sham}},\ }\bibfield  {title} {\bibinfo
  {title} {Stimulated and spontaneous optical generation of electron spin
  coherence in charged gaas quantum dots},\ }\href
  {https://doi.org/10.1103/PhysRevLett.94.227403} {\bibfield  {journal}
  {\bibinfo  {journal} {Phys. Rev. Lett.}\ }\textbf {\bibinfo {volume} {94}},\
  \bibinfo {pages} {227403} (\bibinfo {year} {2005})}\BibitemShut {NoStop}%
\bibitem [{\citenamefont {Wang}\ \emph {et~al.}(2008)\citenamefont {Wang},
  \citenamefont {Li}, \citenamefont {Zhou}, \citenamefont {Kang}, \citenamefont
  {Yun},\ and\ \citenamefont {Gao}}]{Wang2008}%
  \BibitemOpen
  \bibfield  {author} {\bibinfo {author} {\bibfnamefont {C.-L.}\ \bibnamefont
  {Wang}}, \bibinfo {author} {\bibfnamefont {A.-J.}\ \bibnamefont {Li}},
  \bibinfo {author} {\bibfnamefont {X.-Y.}\ \bibnamefont {Zhou}}, \bibinfo
  {author} {\bibfnamefont {Z.-H.}\ \bibnamefont {Kang}}, \bibinfo {author}
  {\bibfnamefont {J.}~\bibnamefont {Yun}},\ and\ \bibinfo {author}
  {\bibfnamefont {J.-Y.}\ \bibnamefont {Gao}},\ }\bibfield  {title} {\bibinfo
  {title} {Investigation of spontaneously generated coherence in dressed states
  of rb85 atoms},\ }\href {https://doi.org/10.1364/ol.33.000687} {\bibfield  {journal} {\bibinfo  {journal}
  {Optics Letters}\ }\textbf {\bibinfo {volume} {33}},\ \bibinfo {pages} {687}
  (\bibinfo {year} {2008})}\BibitemShut {NoStop}%
\bibitem [{\citenamefont {Wang}\ \emph {et~al.}(2009)\citenamefont {Wang},
  \citenamefont {Kang}, \citenamefont {Tian}, \citenamefont {Jiang},\ and\
  \citenamefont {Gao}}]{Wang2009}%
  \BibitemOpen
  \bibfield  {author} {\bibinfo {author} {\bibfnamefont {C.-L.}\ \bibnamefont
  {Wang}}, \bibinfo {author} {\bibfnamefont {Z.-H.}\ \bibnamefont {Kang}},
  \bibinfo {author} {\bibfnamefont {S.-C.}\ \bibnamefont {Tian}}, \bibinfo
  {author} {\bibfnamefont {Y.}~\bibnamefont {Jiang}},\ and\ \bibinfo {author}
  {\bibfnamefont {J.-Y.}\ \bibnamefont {Gao}},\ }\bibfield  {title} {\bibinfo
  {title} {Effect of spontaneously generated coherence on absorption in a
  v-type system: Investigation in dressed states},\ }\href
  {https://doi.org/10.1103/PhysRevA.79.043810} {\bibfield  {journal} {\bibinfo
  {journal} {Phys. Rev. A}\ }\textbf {\bibinfo {volume} {79}},\ \bibinfo
  {pages} {043810} (\bibinfo {year} {2009})}\BibitemShut {NoStop}%
  \bibitem{He2008} Y. He,  et al., Dynamically Controlled Resonance Fluorescence Spectra from a Doubly Dressed Single InGaAs Quantum Dot, Phys. Rev. Lett {\bf 114}, 097402 (2015).
\bibitem [{\citenamefont {Agarwal}(2000)}]{Agarwal2000}%
  \BibitemOpen
  \bibfield  {author} {\bibinfo {author} {\bibfnamefont {G.~S.}\ \bibnamefont
  {Agarwal}},\ }\bibfield  {title} {\bibinfo {title} {Anisotropic
  vacuum-induced interference in decay channels},\ }\href
  {https://doi.org/10.1103/PhysRevLett.84.5500} {\bibfield  {journal} {\bibinfo
   {journal} {Phys. Rev. Lett.}\ }\textbf {\bibinfo {volume} {84}},\ \bibinfo
  {pages} {5500} (\bibinfo {year} {2000})}\BibitemShut {NoStop}%
\bibitem [{\citenamefont {Yang}\ \emph {et~al.}(2008)\citenamefont {Yang},
  \citenamefont {Xu}, \citenamefont {Chen},\ and\ \citenamefont
  {Zhu}}]{YangPRL08}%
  \BibitemOpen
  \bibfield  {author} {\bibinfo {author} {\bibfnamefont {Y.}~\bibnamefont
  {Yang}}, \bibinfo {author} {\bibfnamefont {J.}~\bibnamefont {Xu}}, \bibinfo
  {author} {\bibfnamefont {H.}~\bibnamefont {Chen}},\ and\ \bibinfo {author}
  {\bibfnamefont {S.}~\bibnamefont {Zhu}},\ }\bibfield  {title} {\bibinfo
  {title} {Quantum interference enhancement with left-handed materials},\
  }\href {https://doi.org/10.1103/PhysRevLett.100.043601} {\bibfield  {journal}
  {\bibinfo  {journal} {Phys. Rev. Lett.}\ }\textbf {\bibinfo {volume} {100}},\
  \bibinfo {pages} {043601} (\bibinfo {year} {2008})}\BibitemShut {NoStop}%
\bibitem [{\citenamefont {Yannopapas}\ \emph {et~al.}(2009)\citenamefont
  {Yannopapas}, \citenamefont {Paspalakis},\ and\ \citenamefont
  {Vitanov}}]{Yannopapas}%
  \BibitemOpen
  \bibfield  {author} {\bibinfo {author} {\bibfnamefont {V.}~\bibnamefont
  {Yannopapas}}, \bibinfo {author} {\bibfnamefont {E.}~\bibnamefont
  {Paspalakis}},\ and\ \bibinfo {author} {\bibfnamefont {N.~V.}\ \bibnamefont
  {Vitanov}},\ }\bibfield  {title} {\bibinfo {title} {Plasmon-induced
  enhancement of quantum interference near metallic nanostructures},\ }\href
  {https://doi.org/10.1103/PhysRevLett.103.063602} {\bibfield  {journal}
  {\bibinfo  {journal} {Phys. Rev. Lett.}\ }\textbf {\bibinfo {volume} {103}},\
  \bibinfo {pages} {063602} (\bibinfo {year} {2009})}\BibitemShut {NoStop}%
\bibitem [{\citenamefont {Jha}\ \emph {et~al.}(2015)\citenamefont {Jha},
  \citenamefont {Ni}, \citenamefont {Wu}, \citenamefont {Wang},\ and\
  \citenamefont {Zhang}}]{Jha2015}%
  \BibitemOpen
  \bibfield  {author} {\bibinfo {author} {\bibfnamefont {P.~K.}\ \bibnamefont
  {Jha}}, \bibinfo {author} {\bibfnamefont {X.}~\bibnamefont {Ni}}, \bibinfo
  {author} {\bibfnamefont {C.}~\bibnamefont {Wu}}, \bibinfo {author}
  {\bibfnamefont {Y.}~\bibnamefont {Wang}},\ and\ \bibinfo {author}
  {\bibfnamefont {X.}~\bibnamefont {Zhang}},\ }\bibfield  {title} {\bibinfo
  {title} {Metasurface-enabled remote quantum interference},\ }\href
  {https://doi.org/10.1103/PhysRevLett.115.025501} {\bibfield  {journal}
  {\bibinfo  {journal} {Phys. Rev. Lett.}\ }\textbf {\bibinfo {volume} {115}},\
  \bibinfo {pages} {025501} (\bibinfo {year} {2015})}\BibitemShut {NoStop}%
\bibitem [{\citenamefont {Hughes}\ and\ \citenamefont
  {Agarwal}(2017)}]{Hughes}%
  \BibitemOpen
  \bibfield  {author} {\bibinfo {author} {\bibfnamefont {S.}~\bibnamefont
  {Hughes}}\ and\ \bibinfo {author} {\bibfnamefont {G.~S.}\ \bibnamefont
  {Agarwal}},\ }\bibfield  {title} {\bibinfo {title} {Anisotropy-induced
  quantum interference and population trapping between orthogonal quantum dot
  exciton states in semiconductor cavity systems},\ }\href
  {https://doi.org/10.1103/PhysRevLett.118.063601} {\bibfield  {journal}
  {\bibinfo  {journal} {Phys. Rev. Lett.}\ }\textbf {\bibinfo {volume} {118}},\
  \bibinfo {pages} {063601} (\bibinfo {year} {2017})}\BibitemShut {NoStop}%
\bibitem [{\citenamefont {Karanikolas}\ and\ \citenamefont
  {Paspalakis}(2018)}]{Karanikolas}%
  \BibitemOpen
  \bibfield  {author} {\bibinfo {author} {\bibfnamefont {V.}~\bibnamefont
  {Karanikolas}}\ and\ \bibinfo {author} {\bibfnamefont {E.}~\bibnamefont
  {Paspalakis}},\ }\bibfield  {title} {\bibinfo {title} {Plasmon-induced
  quantum interference near carbon nanostructures},\ }\href {https://doi.org/10.1021/acs.jpcc.8b02703} {\bibfield
  {journal} {\bibinfo  {journal} {J. Phys. Chem. C}\ }\textbf {\bibinfo
  {volume} {122}},\ \bibinfo {pages} {14788} (\bibinfo {year}
  {2018})}\BibitemShut {NoStop}%

 \end{thebibliography}
\end{document}